\documentclass[fleqn,usenatbib]{mnras}

\usepackage{newtxtext,newtxmath, url}

\usepackage[T1]{fontenc}

\DeclareRobustCommand{\VAN}[3]{#2}
\let\VANthebibliography\thebibliography
\def\thebibliography{\DeclareRobustCommand{\VAN}[3]{##3}\VANthebibliography}

\usepackage{graphicx}	
\usepackage{amsmath}	
\usepackage{pdflscape}
\usepackage{multirow}

\title[]{QPEs as Lense-Thirring precession of super-Eddington flows}

\author[M. Middleton et al.]{M. Middleton$^{1}$, A. G\'urpide$^{1}$, T. M. Kwan$^{2}$, L. Dai$^{2}$, R. Arcodia$^{3}$, J. Chakraborty$^{3}$, T. Dauser$^{4}$, \newauthor P. C. Fragile$^{5}$ , A. Ingram$^{6}$, G. Miniutti$^{7}$, C. Pinto$^{8}$ \& P. Kosec$^{3}$
\\
$^{1}$School of Physics \& Astronomy, University of Southampton, Southampton, Southampton SO17 1BJ, UK\\
$^{2}$Department of Physics, The University of Hong Kong, Pokfulam Road, Hong Kong\\
$^{3}$Center for Astrophysics, Harvard \& Smithsonian, Cambridge, MA, USA\\
$^{4}$Dr. Karl Remeis-Observatory and Erlangen Centre for Astroparticle Physics, Sternwartstr. 7, 96049 Bamberg, Germany \\
$^{5}$Department of Physics \& Astronomy, College of Charleston, Charleston, SC, USA\\
$^{6}$School of Mathematics, Statistics and Physics, Newcastle University, Herschel Building, Newcastle upon Tyne, NE1 7RU, UK\\
$^{7}$Centro de Astrobiología (CAB), CSIC-INTA, Camino Bajo del Castillo s/n, 28692 Villanueva de la Cañada, Madrid, Spain\\
$^{8}$INAF, IASF, Palermo, Italy}

\date{Accepted XXX. Received YYY; in original form ZZZ}

\pubyear{2022}

\begin{document}
\label{firstpage}
\pagerange{\pageref{firstpage}--\pageref{lastpage}}
\maketitle

\begin{abstract}
Quasi-periodic eruptions (QPEs) are a recently identified class of X-ray transient associated with tidal disruption events by supermassive black holes, and for which there are multiple possible explanations. In this paper we present a simple model which requires the black hole be spinning, be misaligned with the accretion flow (both conditions of which are almost certainly met) and that the accretion rate is a few times the Eddington limit. We speculate that the resulting Lense-Thirring torques force the disc and entrained outflows to precess, leading to increased X-ray flux when the wind-cone is oriented at lower inclinations to the observer. We test the range of parameters for which this model could explain the period and brightness of the QPE events discovered thus far, and make qualitative comparisons between the observed X-ray spectra and lightcurves to those extracted from GR-RMHD simulations. Overall, we find some areas of promising concordance, and identify challenges related to the details of current simulations. 

\end{abstract}




\section{Introduction}

Quasi-periodic eruptions (QPEs) have been identified in a small (up to nine at the time of writing) but growing number of accreting SMBHs, from their highly characteristic X-ray lightcurves. These are exemplified by an almost constant count rate, showing little variability but with quasi-periodic increases in brightness (hereafter referred to as `events') by a factor of 10-100 depending on energy band. The recurrence timescales of the events are typically $\sim$ hours-days (e.g. \citealt{Giustini2020, Miniutti2023_GSNa, Nicholl2024}) although the recurrence times are observed to vary for some sources (e.g. \citealt{Miniutti2023_GSNa}). 

Spectrally, QPEs are thermal at all phases, with some small fraction of power-law-like emission (similar to some high accretion rate narrow line Seyfert 1 AGN, e.g. \citealt{MD2010, Jin2012}). Over the course of the event, the thermal emission becomes stronger, with an increase in its peak temperature (as far as a characteristic temperature can be obtained) with brightness, although the temperature and peak luminosity of events can change in a given source (see \citealt{Arcordia2022_ero1behaviour, Chakraborty_2024}). The profile of the events is also asymmetric in some cases (e.g. \citealt{Miniutti2019Nature_2MASJ, Giustini2020,  Arcodia2021Nature_ero1&2discovery, Arcodia2024_ero3&4}) with the hard emission rising later and faster than the soft emission, and peaking earlier. Estimates indicate SMBH masses for QPE sources lie in the range of 10$^{5} - 10^{7}$ M$_{\odot}$ (\citealt{Wevers2022, Arcodia2024_ero3&4}) and -- at least in one case -- bolometric luminosities which are close to Eddington (\citealt{Miniutti2023_GSNb}). 

\begin{figure*}
    \centering
    \includegraphics[trim=100 0 100 0, clip, width=0.98\textwidth] {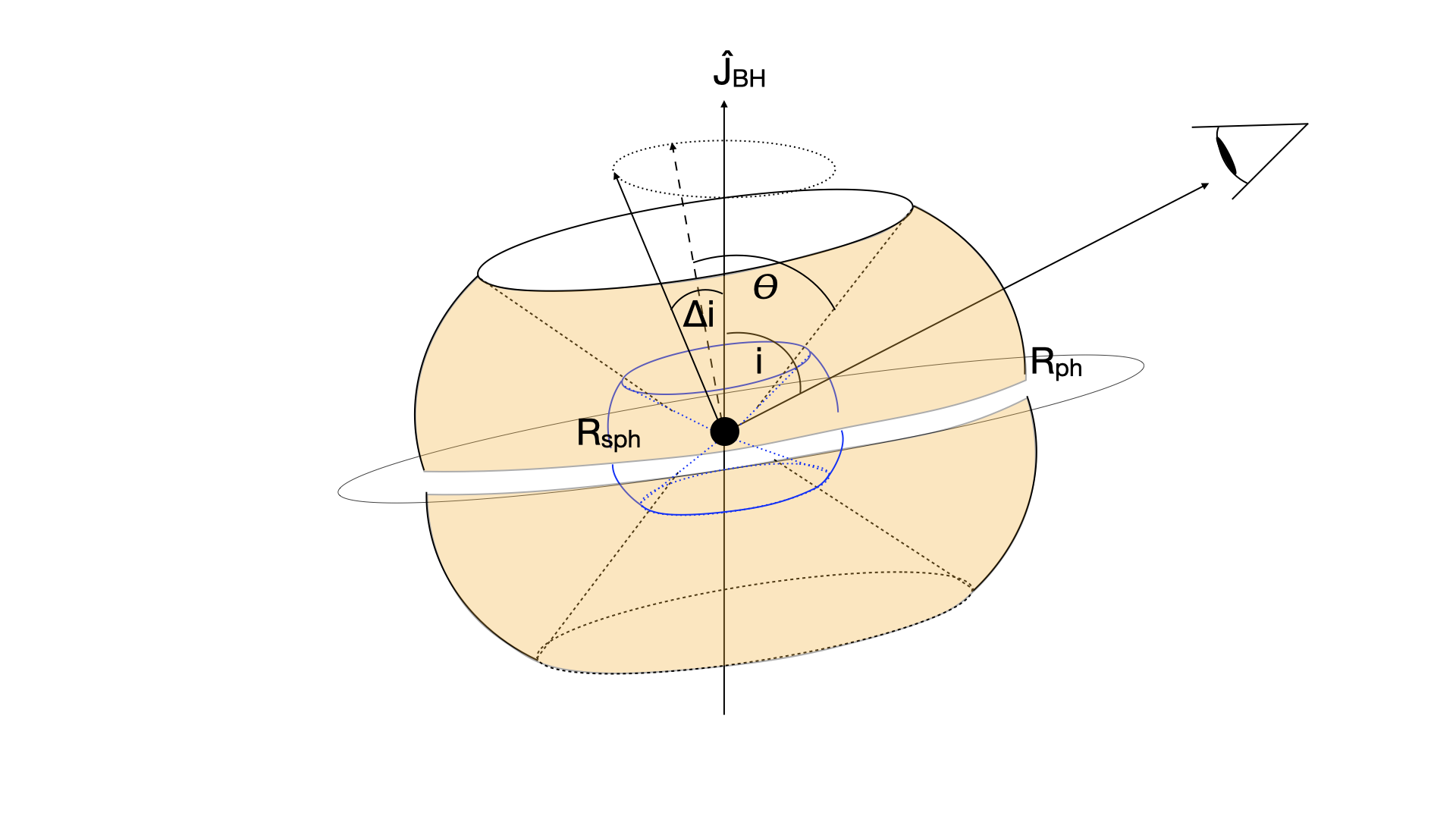}
    \caption{Schematic of our model for a super-Eddington flow in QPE sources. The flow precesses about the black hole angular momentum vector, $\hat{\rm J}_{\rm  BH}$ (with angles indicated and labelled to match our lightcurve analysis (Section 3.3). The arrangement shows the disc (in blue, which inflates within $\sim R_{\rm sph}$) and wind (in orange and radiatively supported out to $\sim R_{\rm ph}$) precessing in-phase; should this not be the case, there will be a lag between the viewing angle to the evacuated wind-cone and the orientation of the inner disc, which we speculate could lead to complex spectral-timing behaviours. Clearly, at large observer inclinations relative to the black hole spin axis ($i$), and for a compact disc (as expected in TDEs and suggested for QPE discs: \citealt{Miniutti2023_GSNa}), the underside, precessing wind cone may also be visible and the emission from this will be typically less bright than from the topside.}
    \label{fig:schem}
\end{figure*}

Various models have been proposed to explain the above phenomenology including the `swiss cheese' rotating wind model (\citealt{Middleton_Ingram2015}), self-lensing of binary SMBHs (\citealt{Ingram2021}), radiative disc instabilities (e.g. \citealt{Raj2021, Kaur2023, Pan2023}) and interactions (tidal interactions or disc crossings) with a secondary, lower mass object, often collectively referred to as extreme mass ratio inspiral -- EMRI -- models (\citealt{Dai10, Dai13, Xian2021, Zhou2022, Wang2022, Metzger2022, Krolik2022, Linial_Sari2023, Linial_Metzger2023, Lu2023, Franchini2023, Tagawa2023, King2023, Linial2024arXiv, Yao2024, Zhou2024}). In this paper we suggest an alternative model which can describe some key phenomenology displayed by QPEs: Lense Thirring precession (LTP) of a super-Eddington accretion disc/wind. 

\section{the model}
The model of super-Eddington LTP was first developed for the case of ultraluminous X-ray sources (ULXs: see \citealt{King2023_ULX} for a review). In ULXs, we can be confident that the flow is super-Eddington, which leads to an increase in the disc scale-height and the powering of winds which are optically thick where the accretion rate is $\gtrsim$ a few times Eddington (\citealt{SS73, P07}). Where the accretion flow is vertically misaligned from the compact object equatorial axis, the disc and winds are proposed to be subject to Lense-Thirring torques and precess as a fluid body at a timescale set by the spin of the compact object and either the position of the spherisation radius in the disc, or the outer photospheric radius of the radiatively driven wind (\citealt{Middleton2018, Middleton2019}). Such precession has recently been observed to occur within numerical simulations (\citealt{Asahina2024}). With regards the spherisation radius, we assume this to be around where the scale-height substantially exceeds the dimensionless viscosity ($\alpha$) within the disc as it inflates due to being locally Eddington. In the case of the outer photospheric radius of the wind, we assume this to coincide with the radial limit at which radiation pressure dominates the outflow (see \citealt{Middleton2019} for details).

As the disc/wind cone tilts relative to our line-of-sight, the spectrum changes in a predictable manner (\citealt{Middleton2015_spectraltiming}) becoming harder and brighter for lower inclination angles (e.g. \citealt{P07, Dauser2017}). In the case of ULXs -- specifically where the compact object is a neutron star -- there are additional torques one must consider (tidal, magnetic) as well as free precession of the neutron star. Conversely, in the case of QPEs, the compact object is {\it known} to be a SMBH and, in the absence of a secondary object (although see e.g \citealt{Franchini2023}), the only likely competing precession/warping mechanism would be that of a radiative warp (\citealt{Pringle1996}). 

It is also well-documented that AGN are likely to be misaligned accretors (e.g. \citealt{Clarke1998, Nagar1999, Kinney2000, Middleton2016}). Moreover, as Tidal Disruption Events (TDEs) are now firmly associated with QPEs (explicitly: \citealt{Sheng2021, Chakraborty2021, Quintin2023, Miniutti2023_GSNa, Nicholl2024, Bykov2024}, or suggested based on a decaying continuum: \citealt{Arcodia2024_ero3&4}),  the mass will be provided without any sort of preferred alignment. This makes Lense-Thirring torques somewhat inevitable (see also {\citealt{Franchini2023} where this is a ingredient in the EMRI models for some -- but not all -- QPEs). 
Precession can then be sustained and is made increasingly likely where the disc is compact (due to short sound crossing times and a lack of viscous damping at large radius: \citealt{Bollimpalli2024MNRAS}). Indeed, QPE discs have been inferred to be extremely compact, with outer radii $\lesssim$ 100s of gravitational radii ({\citealt{Miniutti2023_GSNb, Nicholl2024}}), which might explain the lack of features typically associated with AGN, e.g. the lack of broad lines (\citealt{Wevers2022}) and torus (\citealt{Miniutti2019Nature_2MASJ}). We present the basic schematic of our model in Fig~\ref{fig:schem}.
Interestingly, for TDEs, a viewing-angle-dependent model has also been proposed based on numerical simulations of super-Eddington accretion discs and winds around SMBHs, where the spectrum becomes X-ray dominated when viewing the discs at low inclinations and is otherwise UV/optically dominant at high inclinations \citep{Dai2018, Thomsen.2022, Dai21review}. Although disc precession has not yet been explored in TDE simulations -- although has been invoked as an explanation for the presence of quasiperiodic oscillations (QPOs) during such events (e.g. \citealt{Pasham2019, Pasham2024}), and has been considered analytically (e.g. \citealt{Stone2012, Franchini2016, Zanazzi2019, Teboul2023}) -- one would clearly expect that LTP would lead to more/harder X-ray fluxes as the disc precesses in a periodic fashion.



\begin{figure*}
    \centering
    \includegraphics[width=0.95\textwidth]{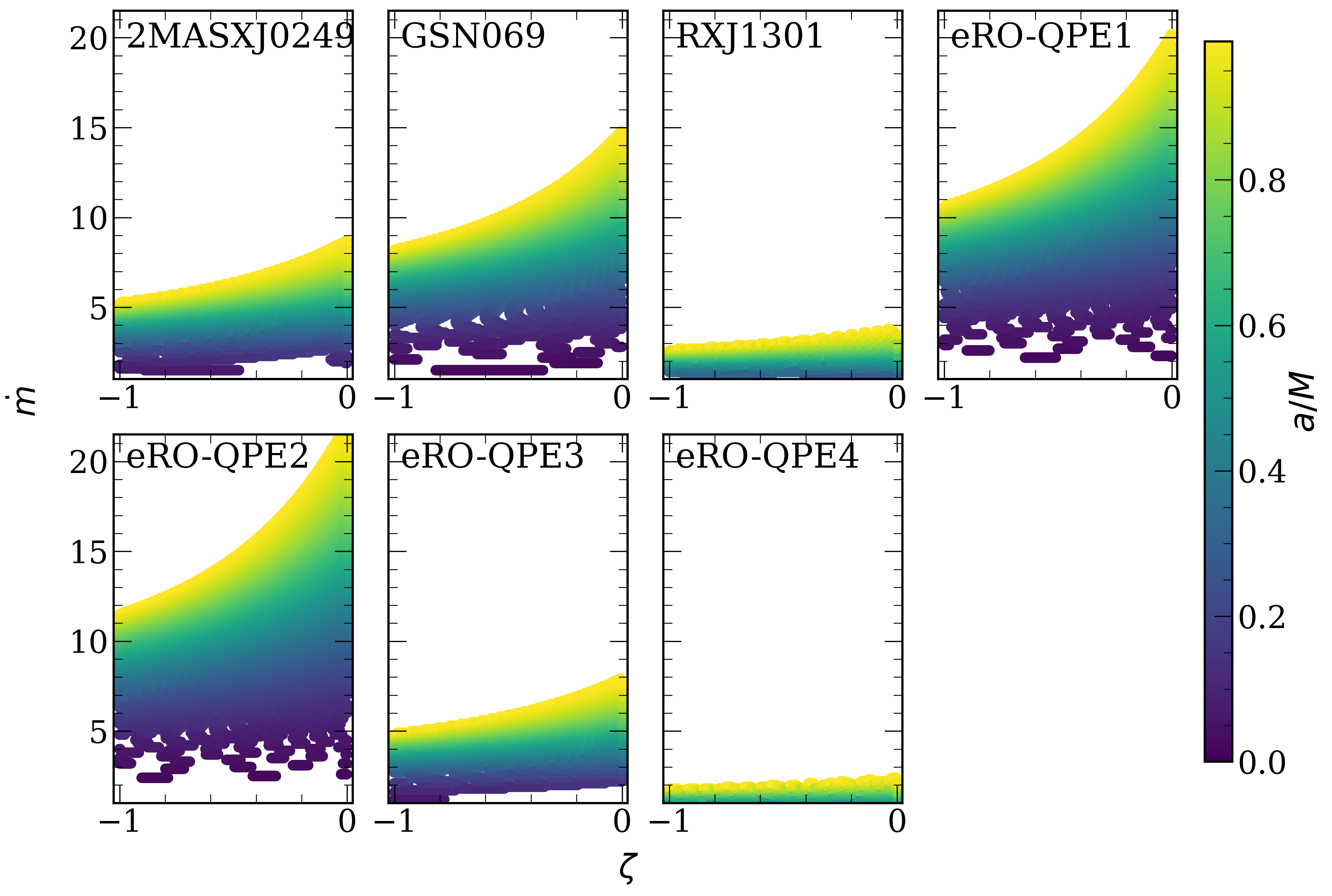}
    \caption{Allowed parameter range for the dimensionless spin ($a/M$), surface density profile index ($\zeta$), and Eddington-scaled accretion rate, $\dot{m}$ when the Lense-Thirring precession period is within 1\% of the assumed QPE period (Table 1). For this search, we set $r_{\rm o} = r_{\rm sph}$ (\citealt{P07}). }
    \label{fig:model1}
\end{figure*}

\begin{figure*}
    \centering
    \includegraphics[width=0.95\textwidth]{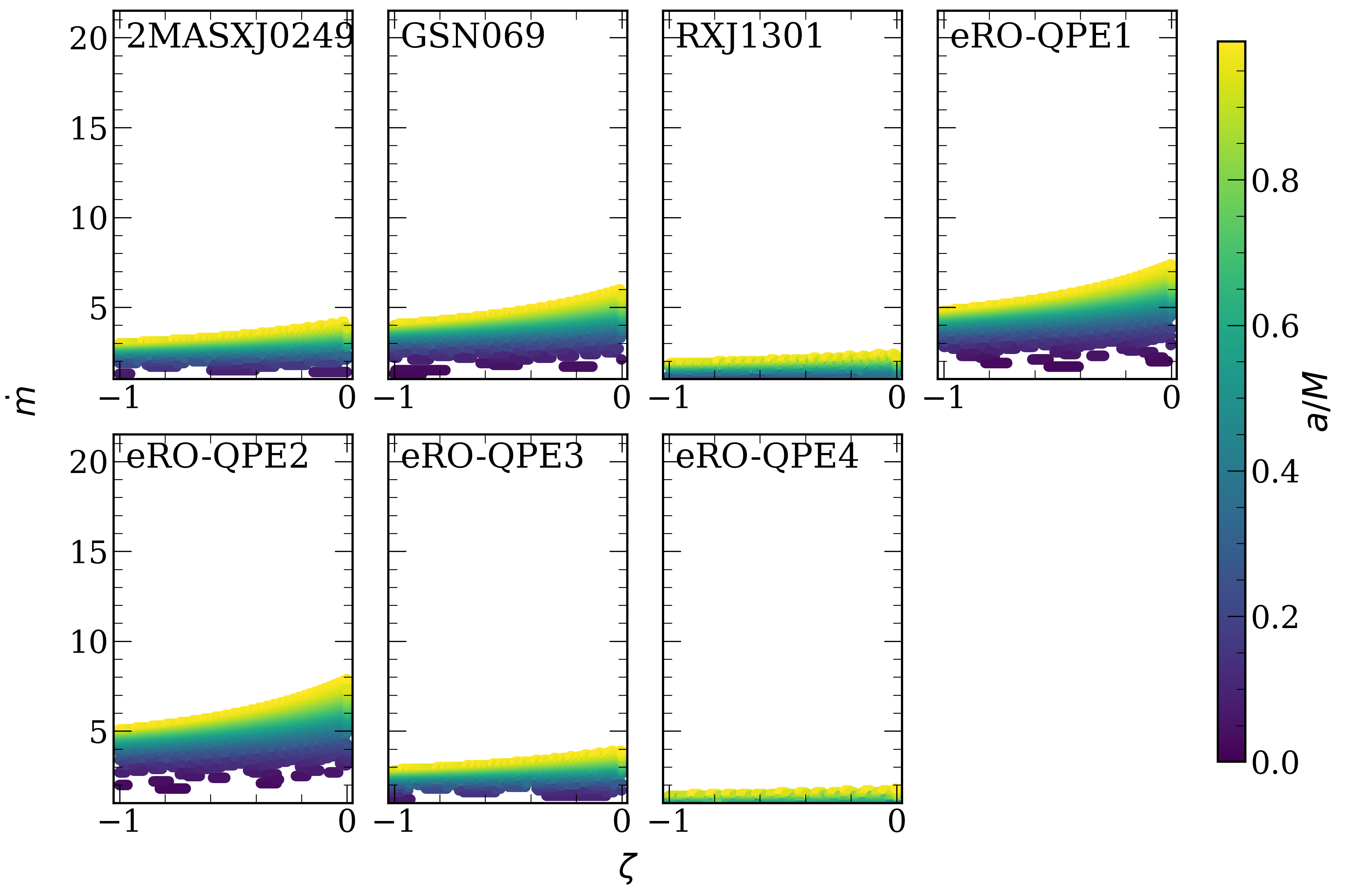}
    \caption{Allowed parameter range for the dimensionless spin (a/M), surface density profile index ($\zeta$), and Eddington-scaled accretion rate, $\dot{m}$ when the Lense-Thirring precession period is within 1\% of the assumed QPE period (Table 1). For this search, we set $r_{\rm o} = r_{\rm ph}$ (\citealt{P07}).}
    \label{fig:model2}
\end{figure*}

In \cite{Middleton2019}, the formula for precession of super-Eddington flows in ULXs was determined based on an analytical radial surface density profile ($\Sigma \propto r^{-\zeta}$) for both the wind and disc (with $\zeta$ = -0.5). Given that AGN are more complex due to the ionisation state of the material (which can affect the surface density profile due to line-driven winds), we instead revert to the original and more general formula from \cite{Fragile2007}:

\begin{equation}
 P_{\rm prec} = \frac{GM}{c^{3}}\frac{\pi \left(1+2\zeta\right)r_{\rm o}^{5/2 - \zeta}}{5-2\zeta}  \frac{r_{\rm i}^{1/2 + \zeta}\left[1-\left(\frac{r_{\rm i}}{r_{\rm o}}\right)^{5/2-\zeta}\right]}{a\left[1-\left(\frac{r_{\rm i}}{r_{\rm o}}\right)^{1/2 + \zeta}\right]}
\end{equation}

\noindent where $a$ is the spin of the black hole in units of the SMBH mass $M$, $r_{\rm i}$ is the inner radius of the disc and $r_{\rm o}$ is the outer radius, both in units of gravitational radius ($R_g\equiv G M/c^2$). In the case of super-Eddington flows, $r_{\rm o}$ is either the spherisation radius in the disc $r_{\rm sph} \approx \dot{m}r_{\rm isco}$ (where $\dot{m}$ is the accretion rate at large radius in units of Eddington -- see \citealt{P07} for details, and $r_{\rm isco}$ is the ISCO radius), or the outer photopsheric radius of the wind, $r_{\rm ph} \approx \dot{m}^{3/2}r_{\rm isco}$ (assuming the additional factor associated with the wind-launching from \citealt{P07} is approximately unity).



\section{Comparison to the QPEs}

\subsection{The time-domain}

We take the estimates for the key parameters (SMBH mass and QPE recurrence time) for the QPE systems shown in Table 1, and invert Equation 1 to infer the dimensionless spin ($a/M$) as a function of Eddington-scaled accretion rate ($\dot{m}$) for a given $\zeta$. We plot the results in Fig~\ref{fig:model1}, assuming $r_{\rm o} = r_{\rm sph}$ and restricting $\zeta$ to take values between 0 and -1.
It is apparent that in all cases (noting that the recurrence period for RXJ1301 and eRO-QPE4 are highly variable -- see Discussion for how this is accommodated within our model), $\dot{m}$ lies between 1-20 with -- as expected from the dependence in Equation 1 -- a lower value of accretion rate requiring a lower value of prograde spin to match the recurrence time. We repeat the analysis assuming $r_{\rm o} = r_{\rm ph}$ and observe that, for the cases of GSN069, eRO-QPE1 and eRO-QPE2, the implied $\dot{m}$ is above a factor of a few (Fig~\ref{fig:model2}), implying that, in these QPE systems, we should be able to most readily detect optically thick outflows. As we discuss later with reference to the quiescent luminosity, we would not expect to observe X-ray luminosities that are 1-20 $\times$ L$_{\rm Edd}$ from QPEs; the intrinsic radiative luminosity generated within the entire flow is at most L$_{\rm Edd}(1+\ln\dot{m})$ which is reduced by advection, a factor to account for energy lost in driving the wind (see \citealt{P07} for details), and, observationally, is further heavily affected by collimation and inclination (see e.g. \citealt{Dauser2017}).

To explore whether our model is able to reproduce the observed luminosities of the QPEs to-date (both within the event and the quiescent bolometric luminosity reported by e.g. \citealt{Miniutti2023_GSNb}), we require the band-limited luminosities at various inclinations to a super-Eddington accretion flow. \cite{Thomsen.2022} carried out three general relativistic radiation magnetohydrodynamic (GR-RMHD) simulations of super-Eddington accretion flows around SMBHs using the \texttt{HARMRAD} code \citep{McKinney.2014}. The SMBH is assumed to have a mass of 10$^{6}M_{\odot}$ and a dimensionless spin parameter of $a/M = 0.8$. The simulation box has an outer boundary at $8500~R_g$. The three simulations achieved a quasi-steady state (inflow/outflow equilibrium) in the inner regions only, with $\dot{m}$ = 7, 12, and 24 respectively (noting that this accretion rate is measured at radii much larger than the ISCO). 
As expected, geometrically and optically thick discs form along the equatorial plane. The discs are magnetically dominated due to the initial simulation set up. The large radiation pressure and magnetic pressure in the discs launch wide-angle winds which move at relativistic speeds (a few $\times$ 0.1c) at low inclinations and slower speeds at high inclinations respectively. As expected, the overall disc and wind gas densities increase at larger mass accretion rates and, in all cases, the density of the wind increases with increasing inclination. In this work, we also carry out an additional simulation with the same set up as the runs in \cite{Thomsen.2022}, except with a lower accretion rate of $\dot{m}$ = 2, achieved in the quasi-steady state. We obtain the radial radiation flux (via M1 closure) as measured by an observer at a large distance ($F^r_{\rm rad}$) from the simulations, to infer the equivalent isotropic, bolometric, radiative luminosity,
\begin{equation}
    L_{\rm iso} (r, i) = 4\pi r^2 F^r_{\rm rad} (r, i).
\end{equation}
\noindent where $r$ is the distance from the black hole 
 and $i$ is the inclination angle. Fig \ref{fig:Ltheta} shows the isotropic luminosity $L_{\rm iso}$ of the outflow at $r=4000\ R_g$, as a function of $i$. Although we are unable to obtain well-constrained values at inclinations above $\sim$60$^{\circ}$ due to the inflow/outflow not having achieved a steady state at large radius, we expect the luminosity to eventually reach a fixed value at high inclinations (perhaps around the scale-height of the disc), somewhere below 10\% of Eddington.

It is apparent from Fig \ref{fig:Ltheta} that there are two regimes in which we could create an increase in flux that might match a QPE; either we observe at moderate inclinations and precession allows us to look into the wind cone (noting that, at the lowest inclinations, the luminosity in our simulations is heavily increased due to the presence of a jet which would be absent in a non magnetically-dominated simulation), or we observe at high inclinations, and precession takes us to more moderate effective inclinations. The reported luminosities for QPEs depend on the model being used (e.g. \citealt{Miniutti2023_GSNb} who find a dynamic range in quiescent/out-of-event luminosity of up to a factor 10 depending on the model) but in quiescence, the luminosity is likely sub-Eddington. This would tend to favour a higher-to-moderate range of inclinations across a precession phase, and also offers the prospect of observing the wind-cone from the underside of the disc. 

 As mentioned in the Discussion, our model would predict aperiodicity should the precession period -- set by the location of the outer radius -- change in response to changes in accretion rate at large radii (e.g. driven by viscous processes, or global changes due to mass loss: \citealt{Middleton2022}). In this case, we would require any changes in accretion rate at large radius to take place on less than the accretion timescale and not many times longer than the sound crossing timescale (otherwise precession will be quasi-stable for long periods). Another possibility for generating aperiodicity would be if the flow was relatively less coherent, which is thought to occur where the outer radius of the precessing region approaches the maximum set by alignment torques: {\citealt{Motta2018}}). The most extreme example of this would be the termination of QPEs (e.g. \citealt{Miniutti2023_GSNb}) when the accretion rate becomes sufficiently high, although the loss of precession could also occur through changes in the accretion or sound crossing timescales (see \citealt{Bollimpalli2024MNRAS}); we return to this point in the Discussion. 

\begin{figure}
    \centering
    \includegraphics[width=\columnwidth]{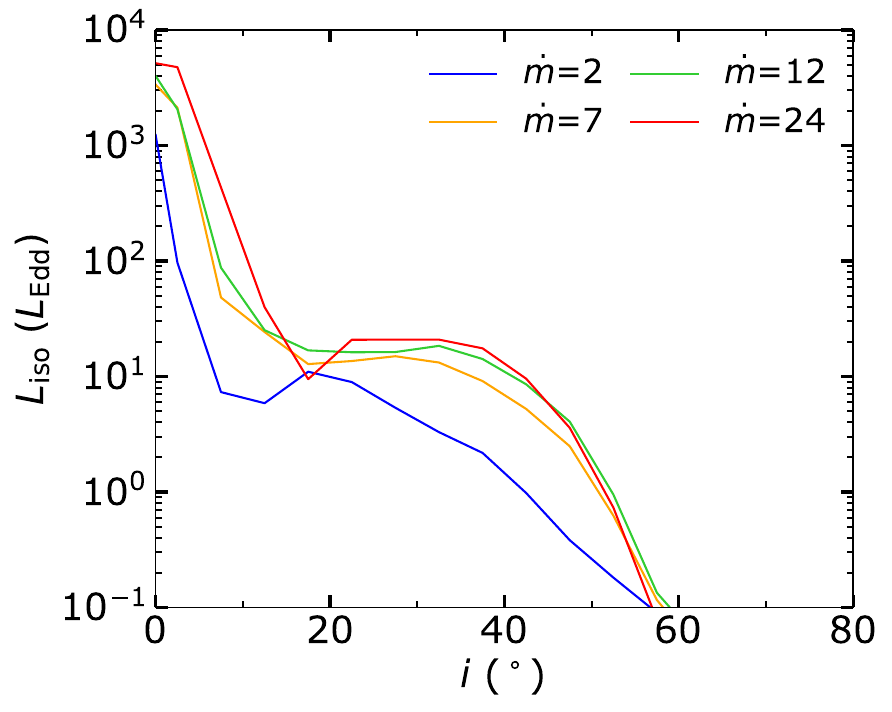}
    \vspace*{-0.4cm}
    \caption{Isotropic, bolometric, radiative luminosity $L_{\rm iso}$ in units of $L_{\rm Edd}$ versus inclination angle ($i$), emerging from the simulations at large distances from the black hole. Blue: $\dot{m}$ = 2, orange: $\dot{m}$ = 7, green: $\dot{m}$ = 12, red: $\dot{m}$ = 24. The shape is qualitatively similar to that obtained numerically by Dauser et al. (2017) and implemented into {\sc ulxlc}, although is higher at the smallest inclinations due to the presence of a jet.}
    
    \label{fig:Ltheta}
\end{figure}
 
 We note the observation of long-short QPE recurrences (\citealt{Miniutti2023_GSNb}), when two events occur closer to one another followed by a longer gap. This can be accommodated within our model due to the presence of two cones of emission; our ability to view these is made possible only when viewed at high inclinations or with large precession angles. However to match observations would also likely require some level of asymmetry above and below the disc to produce signals which are not strictly anti-phase (dealt with within the EMRI models through an elliptical orbit of the compact object through the disc). We note that, in cases where the undercone emission is sufficiently visible, the `true' precession period would be that between brightenings from each cone rather than between events.




\begin{table}
\centering
\begin{tabular}{c|c|c|c}
\hline
Name & log Mass (M$_{\odot}$) 
& Period (ks) & Distance (Mpc)\\
\hline 
GSN069 & 5.99 $\pm$ 0.50 
& 32.4 & 75 \\
RXJ1301 & 6.10 $\pm$ 0.42 
& 16.5 & 99.3\\
eRO-QPE1 & 5.78 $\pm$ 0.55 
& 68.4 
& 215.5  \\
eRO-QPE2 & 4.96 $\pm$ 0.54 
& 8.6 & 72.9\\
eRO-QPE3 & 6.49$^{+0.23}_{-0.54}$ & 72 & 100.4\\
eRO-QPE4 & 7.62$^{+0.21}_{-0.40}$ & 44 &  186.8\\
2MASJ & 5.29 $\pm$ 0.55 
& 9 & 77.5\\

\hline
\end{tabular}
\caption{Estimates for the QPE recurrence times, SMBH masses and distances used in this paper. In the cases of RXJ1301 and eRO-QPE4, we have taken an approximate mean recurrence time but highlight that the recurrence time can change substantially for these sources (see \citealt{Giustini2020} and \citealt{Arcodia2024_ero3&4} for details and the Discussion for an explanation for this within the model). Distances are obtained from the Hubble flow (see text for details).} \label{tab:qpe_values}
\end{table}



\subsection{Energy spectra}

It is clear from our analysis thus far that the LTP timescales could lie within range of observed QPE recurrence times, and precession could potentially explain the rise and fall seen in the events, as well as the approximate luminosities observed (see the Discussion for issues related to the asymmetry of the event both in brightness and energy). We now explore whether QPE energy spectra could also be consistent with precession of a super-Eddington flow, specifically when viewed at high to moderate inclinations.

For our observational analysis, we utilise data from observations taken by {\it XMM-Newton}, corresponding to various phases of QPE lightcurves taken from the literature (\citealt{Giustini2020, Chakraborty2021, Arcodia2024_ero3&4, Miniutti2023_GSNa}). In all cases, the background is taken from the surrounding chip  rather than the source in quiescence; this makes the explicit assumption that, if the quiescent emission is a distinct spectral component, the event will dominate the spectrum. In practice, as we are currently only seeking to broadly characterise the events, this is not an important issue and we will return to this topic in a forthcoming study. In brief, we use the following datasets:

\begin{itemize}
    \item {\bf RXJ1301}

Due to the lack of a well-defined period for the QPE events from this source, we selected phase-resolved EPIC-pn spectra from 15/34 QPE events which were deemed to behave in a similar way (based on similar temperatures and luminosities from fitting a simple {\sc zbbody} model to each individual QPE event). The events were taken from all five {\it XMM-Newton} observations of RX J1301.9+2747 (RXJ1301 hereafter) between 2000 and 2022. We selected five phases (two covering the rise, the peak, and two covering the decay) lasting 4$\sigma$/5 seconds each, where $\sigma$ is the width of the best-fitting Gaussian to the 0.2-2 keV QPE profile. \\

    \item {\bf 2MASJ024916.6-04}

Due to the poorer data quality presently available for 2MASJ024916.6-04 (2MASJ hereafter), we extracted only three phases, the rise (1ks), peak (500s) and decay (1ks) of the event from the EPIC-pn data taken in 2006 (\citealt{Chakraborty2021}).\\ 
    
    \item {\bf GSN069}

We extracted EPIC-pn phase-resolved spectra of GSN069 from the {\it XMM-Newton} observation taken in 2019 (see e.g. Fig 17 in \citealt{Miniutti2023_GSNa}) where we ignored any differences between strong and weak QPEs, i.e. the QPE profile is extracted from a light curve folded on the average separation between events. As with RXJ1301, the high data quality allowed us to resolve the event into five phases, two covering the rise and decay, and one for the peak. \\
    
    \item {\bf eROSITA QPEs}

We extracted EPIC-pn spectra from observations of the QPE sources discovered by {\it eROSITA} (\citealt{Arcodia2021Nature_ero1&2discovery, Arcodia2024_ero3&4}). The data processing is described in the related publications. For eRO-QPE1, the isolated event in the {\it XMM-Newton} observation named `eRO-QPE1 Obs2' in \cite{Arcordia2022_ero1behaviour} was used for phase-resolved spectroscopy. Spectra corresponding to the rise, peak and decay phases include good-time intervals in time bins of $7$, $10$ and $24\,$ks, respectively, and are roughly separated at a count rate of $\sim0.6$ ct/s at the brighter end. For eRO-QPE2, given the lower signal-to-noise of individual events (\citealt{Arcodia2021Nature_ero1&2discovery}), phase-folded spectra were extracted (\citealt{Arcodia2024}) combining all the events of the 2020 {\it XMM-Newton} dataset. For eRO-QPE3 and eRO-QPE4, phase-folded spectra are adopted from Fig A9 of \cite{Arcodia2024_ero3&4}, however, due to limited counts, we are unable to use the first phase (Rise 1, see Tables A1 and A2) for either source.  
    
\end{itemize}

In order to obtain a crude description of the spectrum (for a detailed spectral analysis we direct the reader to {\citealt{Giustini2020, Giustini2024,  Arcodia2021Nature_ero1&2discovery, Arcodia2024_ero3&4, Chakraborty2021, Miniutti2023_GSNb}}), we fit each QPE datset in {\sc xspec} (\citealt{Arnaud1996}) with a thermal continuum model of {\sc tbabs*diskbb} where we use the abundances from \cite{Wilms2000} and the lower limit on the column is set to the Galactic line-of-sight value ({\citealt{HI4PI}}). In the cases of GSN069, eRO-QPE1 and eRO-QPE2, the data are sufficient to require the inclusion of a weak power-law ({\sc po}) which is often poorly constrained; in these cases we freeze the power-law index to 2 (i.e. flat in EF$_{\rm E}$). The best-fit parameters and uncertainties are shown in Tables A1-3 and unfolded spectral data, model and residuals shown in  Fig~\ref{fig:spectra1} and Fig~\ref{fig:spectra2}. In a number of sources -- especially GSN069, RXJ1301 and eRO-QPE1, two of which we have inferred may be higher Eddington rate systems (see Fig~\ref{fig:model1}) and therefore may have higher density winds -- there are structured residuals to the best-fitting model (see also \citealt{Miniutti2023_GSNa}). In ULXs, such residuals have been shown to be well-established indicators of the presence of radiatively driven outflows from a super-Eddington disc (typically with v $\ge$ 0.2c: \citealt{Middleton2014, Middleton2015_winds, Pinto2016, Kosec2021}), this has now been confirmed in the case of GSN069, with RGS data indicating a mixture of narrow absorption lines with projected velocities of $\sim$ 0.01c, and possibly broader lines in the CCD spectrum (\citealt{Kosec2024}). This detection would appear to be consistent with the overall model of super-Eddington accretion, as, given the inclinations we have predicted based on the observed luminosity, we would expect to observe winds which are somewhat  slower than those seen in ULXs and subtend a large covering fraction from our perspective.


\begin{figure*}
    \centering
    \includegraphics[trim=150 100 300 50, clip, width=0.98\textwidth]{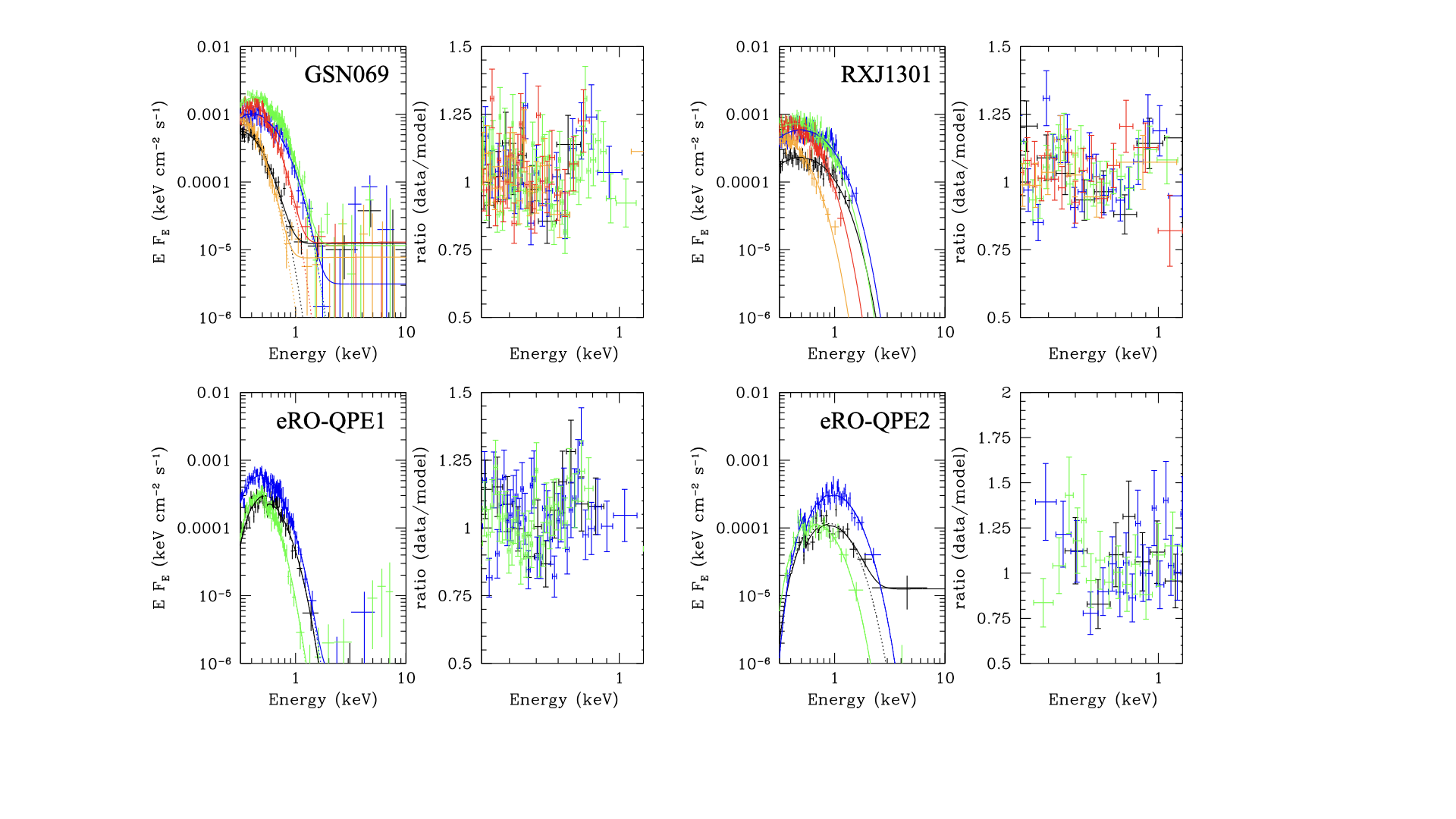}
    \caption{Spectral data and best-fitting models for the known QPE sources to-date. Each source and QPE phase is fitted with a simple absorbed disc blackbody model ({\sc tbabs*diskbb}) and, in those cases where data permits, with an additional power-law (see text for details). The colour scheme in terms of event chronology is black-blue-green or black-blue-green-red-orange where we have greater data quality and phase resolution.
    Residuals to the fits at soft energies are shown in the right hand panels; in a number of cases (especially in GSN069, RXJ1301 and eRO-QPE1) there are indicators of structure which -- in the case of GS069 -- has been resolved into lines associated with a wind, with the RGS-observed wind-phase outflowing at speeds less than that seen in ULXs (\citealt{Middleton2015_winds, Pinto2016}). See Tables A1 and A2 for the best-fitting parameters and associated 1$\sigma$ errors.}
    \label{fig:spectra1}
\end{figure*}

\begin{figure*}
    \centering
    \includegraphics[trim=100 100 300 0, clip, width=0.98\textwidth]{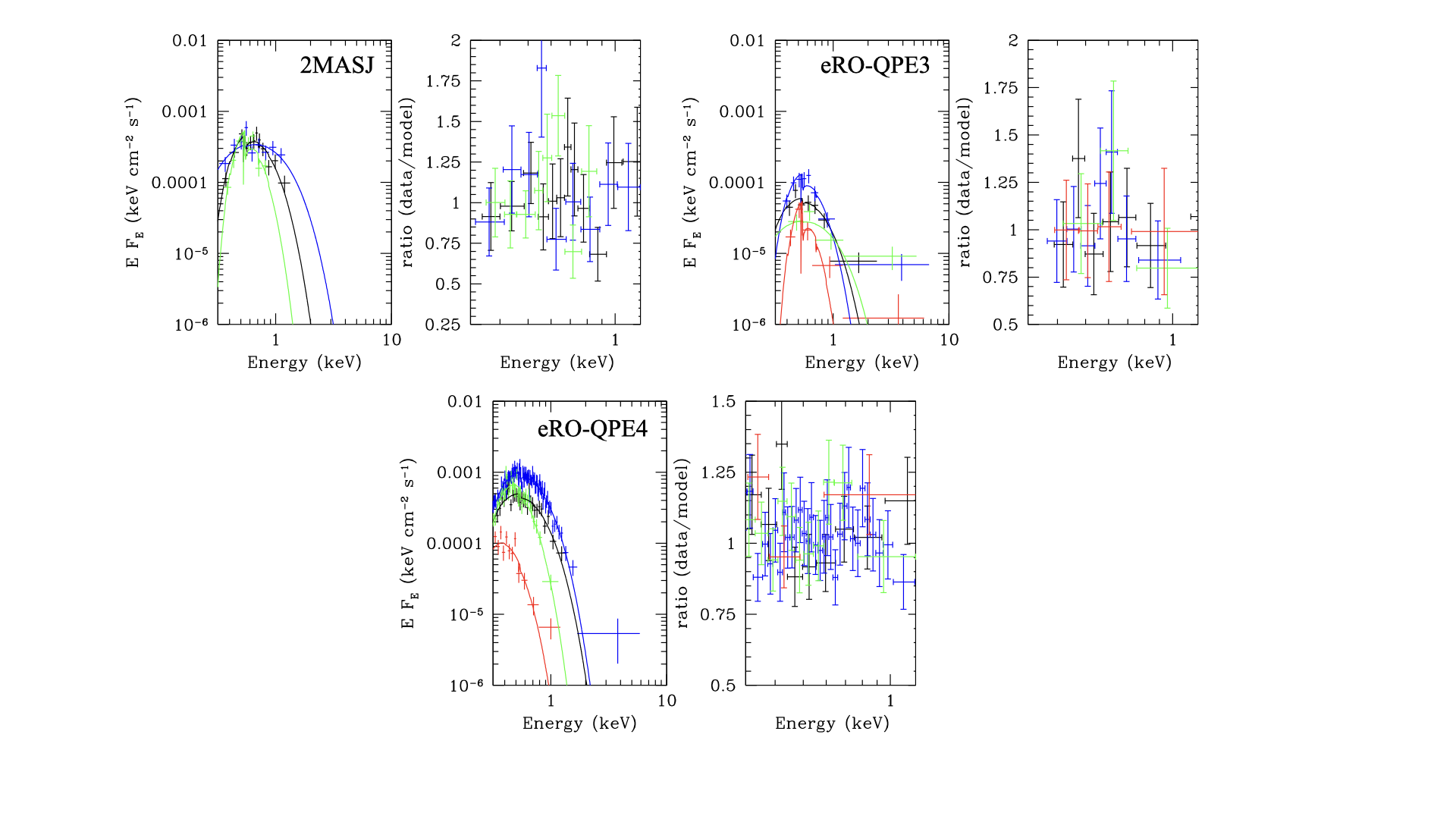}
    \caption{As for Figure 5. In the case of ero-QPE3 and 4, the colour scheme in terms of event chronology is black-blue-green-red. See Tables A2 and A3 for the best-fitting parameters and associated 1$\sigma$ errors.}
    \label{fig:spectra2}
\end{figure*}

We proceed to compare the observed spectra to those extracted from the super-Eddington accretion flow simulations described above (\citealt{Thomsen.2022}). To obtain energy spectra from the simulations, we use the Monte-Carlo radiative transfer code, \texttt{Sedona} \citep{Kasen2006} to post-process the simulated accretion flow and re-calculate the SED of the escaping radiation at different inclination angles within bins: $i_{\rm bin}=$ [0$^\circ$-- 5$^\circ$], [5$^\circ$-- 10$^\circ$], [10$^\circ$-- 15$^\circ$], [15$^\circ$-- 20$^\circ$], ..., [60$^\circ$-- 65$^\circ$]. Within each inclination bin, we first obtain the time-$\theta$-$\phi$-averaged profile of the simulated accretion flow, and then perform the radiative transfer calculations at higher-frequency resolution to calculate the SED more accurately. We follow \cite{Thomsen.2022} and inject an X-ray blackbody peaking at $T=10^6$ K, scaling the escaped bolometric luminosity to the isotropic luminosity $L_{\rm iso}$ from the GR-RMHD simulations (Fig \ref{fig:Ltheta}). The post-processed spectra (noting the caveats below) are shown in Fig~\ref{fig:GRsimspectra}, with a reduced number of inclination bins shown for clarity. It is apparent that there is an overall trend of increasing temperature and X-ray luminosity with decreasing inclination angle, as already invoked for both TDEs (\citealt{Dai2018}) and ULXs (\citealt{P07, Middleton2015_spectraltiming}). 


We note that there can be a few factors leading to the underestimation of the temperature of the emission obtained by the MC simulations: 
1) the starting temperature of $T=10^6\ $K is derived from the radiation temperature seen in the GR-RMHD simulations, which employ only simple radiative transfer physics and use grey opacities. More detailed treatment of the radiation physics, e.g. self-consistent computation of Compton scattering (see {\citealt{Mills2023}}) are known to result in harder MC spectra;  2) the MC radiative transfer calculation conducted is 1D, therefore, it cannot properly address 2D/3D effects, such as bulk Comptonization in the winds (as they flow towards the SMBH and then get squeezed near the turning point before flowing out, \citealt{Kitaki17}) or X-ray photons scattered first in the wind cone and then leaking out at moderate inclinations; 3) when performing the MC calculations, we enforce an escaped luminosity of $L_{\rm Edd}$ and then scale up the spectrum according to $L_{\rm iso}$. One can see from Fig. \ref{fig:Ltheta} that $L_{\rm iso}$ can largely exceed $L_{\rm Edd}$ at low inclinations, which means the photon temperature and luminosity injected are likely lower than the actual values. These factors mostly affect the low-to-moderate inclination MC spectra; whilst the large number of scatterings may tend to erase some of the initial conditions at higher inclinations (so long as the the wind is highly optically thick, i.e. at high super-Eddington accretion rates), we still anticipate that the spectrum will become harder, especially at lower super-Eddington accretion rates. In addition to the above issues regarding the starting temperature, we also note that the input spectrum is injected only from the innermost regions rather than being self-consistently distributed radially and vertically -- this may lead to a further distortion away from our MC calculations.

Being aware of the above issues, we compare peak temperatures and luminosities from the GR-RMHD simulations with those derived from the spectral fitting (see Tables A1, A2 and A3) in Fig~\ref{fig:tracks}. The peak temperature of the SEDs from the simulations were corrected by a factor of 2.82 to obtain kT$_{\rm disc}$. We note that both of these temperatures (from the spectral fitting and from the GR-RMHD simulations) are not colour corrected. The 0.3–10~keV luminosities and their uncertainties were obtained from the best-fitting models to the data by including a {\sc cflux} component in our model (e.g. {\sc tbabs*cflux*diskbb}).
For source distances, we assume Hubble flow and use values for the redshift from the literature (see Tables A1-3 for details), with H$_{\rm 0}$ = 73 km/s/Mpc (\citealt{Riess2022}), $\Omega_{\rm M}$ = 0.286 and $\Omega_{\rm vac}$= 0.714. 
From Fig~\ref{fig:tracks} it is clear that, for our model to work, the QPE sources would need to be viewed at high inclinations, however, the temperatures of the predicted MC spectra appear too low to describe the majority of observations. As we mention above (and revisit in the Discussion), this may be a result of the initial conditions for the MC calculations.

\begin{figure*}
    \centering
    \includegraphics[width=\textwidth]{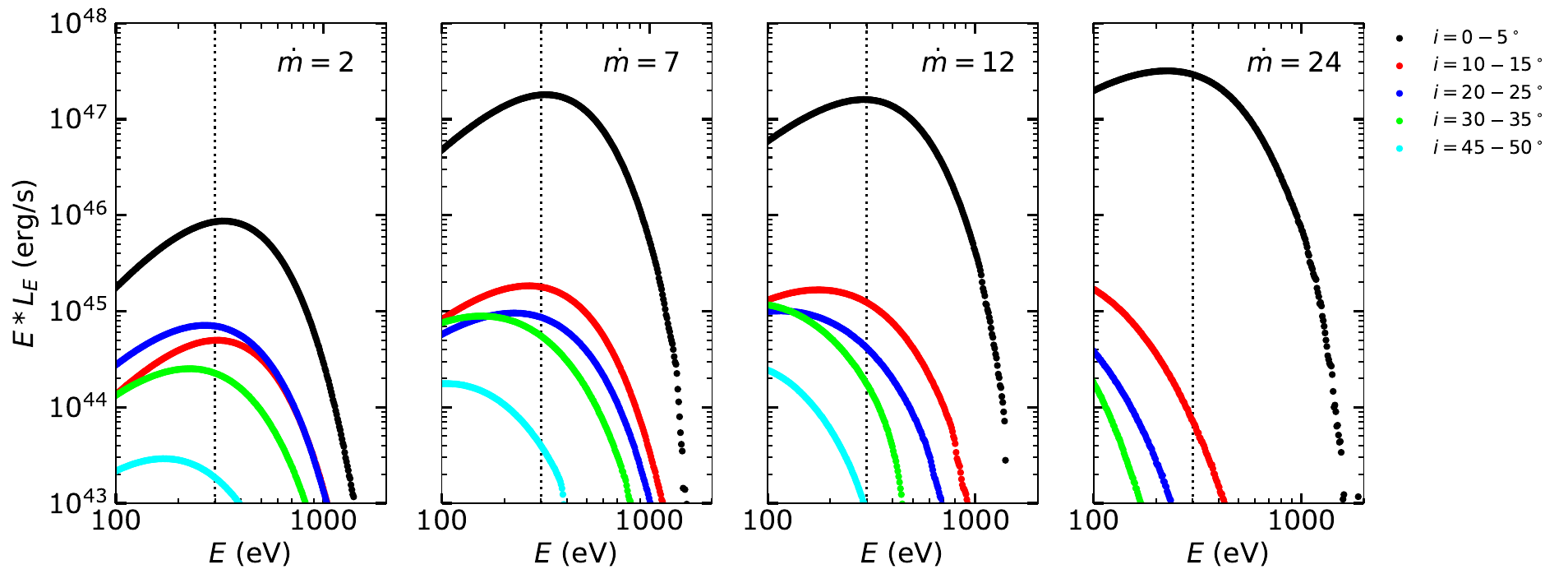}
    \vspace*{-0.4cm}
    \caption{Post-processed spectra from GR-RMHD simulations of $\dot{m}$ = 2, 7, 12 and 24 onto a 10$^{6}$ M$_{\odot}$ SMBH with a/M = 0.8. The colours indicate inclination bins over which the flux is integrated (not all are shown for the sake of clarity), black: 0-5$^{\circ}$, red: 10-15$^{\circ}$, blue: 20-25$^{\circ}$, green: 30-35$^{\circ}$ and cyan: 45-50$^{\circ}$. The vertical dotted line indicates the lower limit of the {\it XMM-Newton} bandpass (0.3 keV) used within our spectral fitting and in Fig~\ref{fig:tracks}.}
    \label{fig:GRsimspectra}
\end{figure*}

\begin{figure*}
    \centering
    \includegraphics[width=\textwidth]{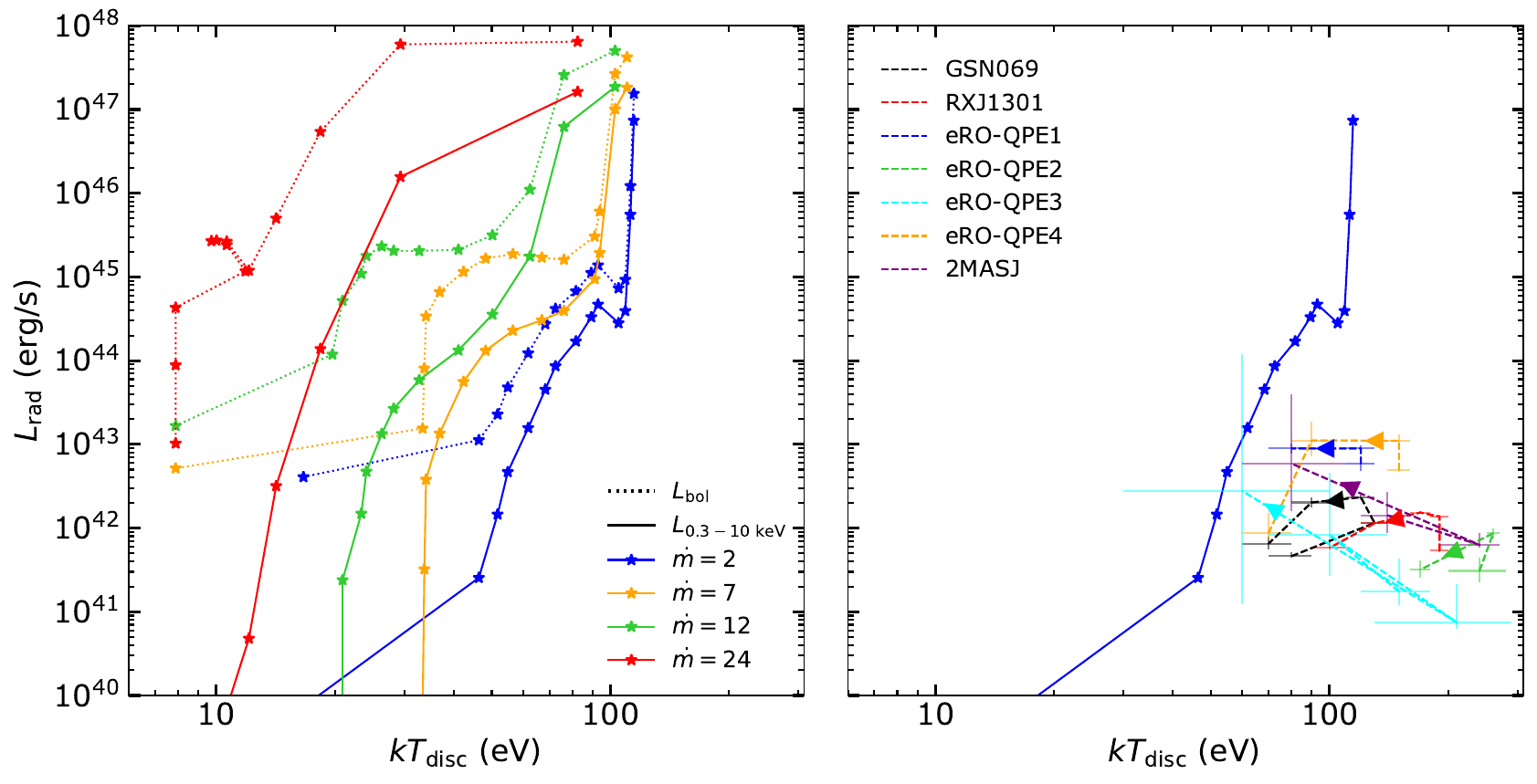}
    \vspace*{-0.4cm}
    \caption{Left hand panel: the dotted lines indicate the bolometric, radiative luminosity $L_{\rm bol}$ from the GR-RMHD simulations of $\dot{m}$ = 2 (blue), 7 (orange), 12 (green) and 24 (red) across the full inclination range of the simulations (with small inclinations having a higher luminosity). The solid lines indicate the 0.3-10 keV radiative luminosity ($L_{\rm 0.3 - 10 \ keV}$) with the same colour scheme. Right-hand panel: the best-fitting {\sc diskbb} temperatures ($kT_{\rm disc}$) versus unabsorbed luminosities for the QPE sources considered in this paper. Black: GSN069, red: RXJ1301, blue: eRO-QPE1, green: eRO-QPE2, cyan: eRO-QPE3, orange: eRO-QPE4 and purple: 2MASJ. Arrows indicate the direction of evolution through the event. The solid line shows the 0.3-10 keV radiative luminosity from the GR-RMHD simulations at $\dot{m}$ = 2 (blue) for reference.} 
    \label{fig:tracks}
\end{figure*}


    




\subsection{Direct lightcurve modelling}\label{sub:lightcurve_modelling}
  



By combining the outputs of the GR-RMHD simulations and a suitable kinematic model, we can attempt to directly (albeit somewhat crudely) model the X-ray lightcurves of QPEs. 

Due to the aperiodicity of many of the QPE sources and an assumed periodic precessional clock (i.e. in the absence of any changes in accretion rate which would drive changes in the QPE recurrence time), we restrict our analysis to eRO-QPE2 and GSN069, as these have shown the most regular and well-defined events. We used two of the {\it XMM-Newton} observations where the QPEs have been clearly detected for these sources ({\sc obsids} 0872390101 and 0831790701 for eRO-QPE2 and GSN069 respectively). To extract the lightcurves, we reprocessed the data using \texttt{epproc} in SAS version 20.0.0. We filtered the lightcurves for particle flaring by first extracting background 10--12 keV lightcurves and then inspected these  visually to set a threshold count-rate to reject times of high-background flaring. We selected {\sc pattern} $\le$ 4 events and used \texttt{eregionanalyse}, with the input source coordinates, to select a suitable source region. The circular regions, as determined by the task, had radii of 24" and 47" for eRO-QPE2 and GSN069 respectively. A $\sim$47" radius circular region on the same chip and as close as possible to the source region, was selected for background lightcurve extraction. We also corrected the lightcurves for effects including losses due to vignetting, chip gaps and bad pixels using \texttt{epiclccorr}. In order to convert the count-rates to approximate Eddington luminosities for the modelling (see below), we used the count-rates of the quiescent, rise, and peak phases, and their derived luminosities (see Tables~\ref{table:a1} and ~\ref{table:a2}) to obtain a crude mapping between count-rate and luminosity. In order to obtain the quiescent luminosity for these QPE sources, we fit the out-of-event data with the {\sc tbabs*cflux*diskbb} model (with an additional power-law where data requires). In the case of eRO-QPE2, this model is a poor description of the quiescent data, indicated by the unabsorbed luminosity being greater than that in the rising phase. To avoid this issue, and in the case of only this source, we tie the neutral column in quiescence to that of the rising phase to obtain the value for the luminosity in this phase only. We subsequently obtain quiescent luminosities of 2.52$^{+0.71}_{-0.54} \times 10^{41}$ erg/s for eRO-QPE2 and 3.70$^{+0.49}_{-0.42} \times 10^{41}$ erg/s for GSN069 respectively. Whilst these numbers are somewhat inaccurate due to the modelling uncertainty, they are sufficient for the proof-of-principle test we are performing.
We note that, when fitting a precession model to the luminosity values we use only the rise and not the decay luminosity and so ignore the known hysteresis in the count rate and luminosity space (\citealt{Arcordia2022_ero1behaviour, Miniutti2023_GSNa}) which we cannot address with this simplified model.


In order to create a model to fit the QPE lighcurves, we have coupled the band-limited GR-RMHD simulations (Section 3.2) with the kinematic model from \cite{Abell1979}, which was derived to explain the precession of the super-Eddington disc and wind in the high-mass X-ray binary, SS433. In this model, the instantaneous inclination of the system with respect to the line-of-sight $\alpha(t)$ at a given time (cf Fig~\ref{fig:schem}), $t$, can be expressed as:
\begin{equation}
    \cos \alpha (t) = \cos \Delta i \cos i + \sin \Delta i \sin i \cos \{2 \pi [(t-t_\mathrm{0}) / P + \phi]\}
\end{equation}
\noindent where $i$ is the inclination angle or line-of-sight angle with respect to the rotational axis, $\Delta i$ is the precessional cone  half opening angle, $P$ is the QPE period and $\phi$ is the phase of the precession cycle at the beginning of the lightcurve (at time $t_\mathrm{0}$). Note that the solutions for $\cos \alpha (t)$ are symmetric with respect to $i$ and $\Delta i$ in the sense that any solution pair ($i$, $\Delta i$) has a reciprocal solution where $\Delta i$ and $i$ are interchanged. Here we present only the set of solutions for which $i > \Delta i$. We note that we do not presently include emission from the underside of the flow (but point out its relevance in the Discussion).

For a given set of $i, \Delta i$, $P$ and $\phi$, we calculate $\alpha (t)$ and convert it to X-ray luminosity by interpolating the mapping between inclination and X-ray luminosity provided by the GR-RMHD simulations (e.g. Figure~\ref{fig:Ltheta}) for a given $\dot{m}$. We note that $\dot{m}$ is restricted to the values used in the GR-RMHD simulations (2, 7, 12 and 24). In practice, we cast the MCMC samples to the closest $\dot{m}$ from the simulations\footnote{See \url{https://github.com/dfm/emcee/issues/150} for a discussion on this aspect.}. The final, best-fitting estimate for $\dot{m}$ is then taken as the average from the posterior samples (see below) and can lie between the discrete $\dot{m}$ values from the GR-RMHD simulations.

As stated in Section 3.1, the exact $L(i)$ dependence at large inclinations ($i \gtrsim$67$^\circ$) is not known from the GR-RMHD simulations due to the flow not having reached steady-state at large radius. As a result of this uncertainty, we have tested two approaches: one where the GR-RMHD outputs were extrapolated down to $i = 90^\circ$ 
using a monotonic cubic spline, and another where we assumed $L=0$ beyond the last inclination angle computed from the GR-RMHD simulations. In both instances, an additional constant term ($A$) is added to our model to account for the approximately steady emission between events (in practice this sets $L=A$ above the cut-off angle). We speculate that the constant flux level might originate either from emission at higher inclinations not covered by the simulations and asymptoting to a fixed luminosity (similar to the emission between 20 and 40 degrees in Figure 7), or alternatively, emission from the non-precessing disc at somewhat larger radii. The best fitting results were obtained with the latter approach, i.e. assuming $L = 0$ (in effect $L = A$), for $i \gtrsim 67^\circ$, and so we present only the results of this analysis as an example of how precession might be a viable explanation for such events. 

The model was first evaluated on a grid 50 times finer than the temporal resolution of the data, and then an average across these bins was taken for the final model-to-data comparison. This more accurate treatment of the observational effects on the model 
eliminated solutions where the flux showed unrealistic variations in flux (e.g. large spikes) in between data points. We then applied the model to the 0.3--10 keV, 300s-binned, background-subtracted EPIC-pn lightcurves of eRO-QPE2 and GSN069 (when showing QPEs), and  performed a $\chi^2$-minimization routine using \texttt{L-BFGS-B} to find a reasonable starting set of parameters for the MCMC sampler. All parameters were free to vary, except for $\dot{m}$, which was allowed to vary between 1 and 26, and the period, $P$, which was only allowed to vary in a range 0.95$P_\mathrm{0} < P < 1.05 P_\mathrm{0}$ around the reported values in Table~\ref{tab:qpe_values}. The range for $\dot{m}$ reflects the range from the numerical simulations, while the range of $P$ was wide enough to allow room to fit for $P$ and account for its uncertainties, but narrow enough to reject solutions deviating substantially from the expected values. After the initial fit, we distributed 200 walkers around the best-fit parameter space by sampling from a Gaussian distribution with a spread equal to 30\% of each of the best-fit parameter values. We then used \texttt{emcee} (\citealt{Foreman-Mackey2013}) to run an MCMC sampler for 200,000 steps. In order to build the posteriors, we discarded the first 5$\tau$ steps and thinned the chains by $\tau$/2, where $\tau$ was the autocorrelation time of the chains. The final posteriors are presented in Figure~\ref{fig:ulxlc_posteriors}, and the constraints on the best-fit parameters are given in Table~\ref{tab:ulxlc_fits} (noting that the errors are underestimates given the unknown systematic uncertainty in the model itself).

\begin{figure*}
    \centering
    \includegraphics[width=0.49\textwidth]{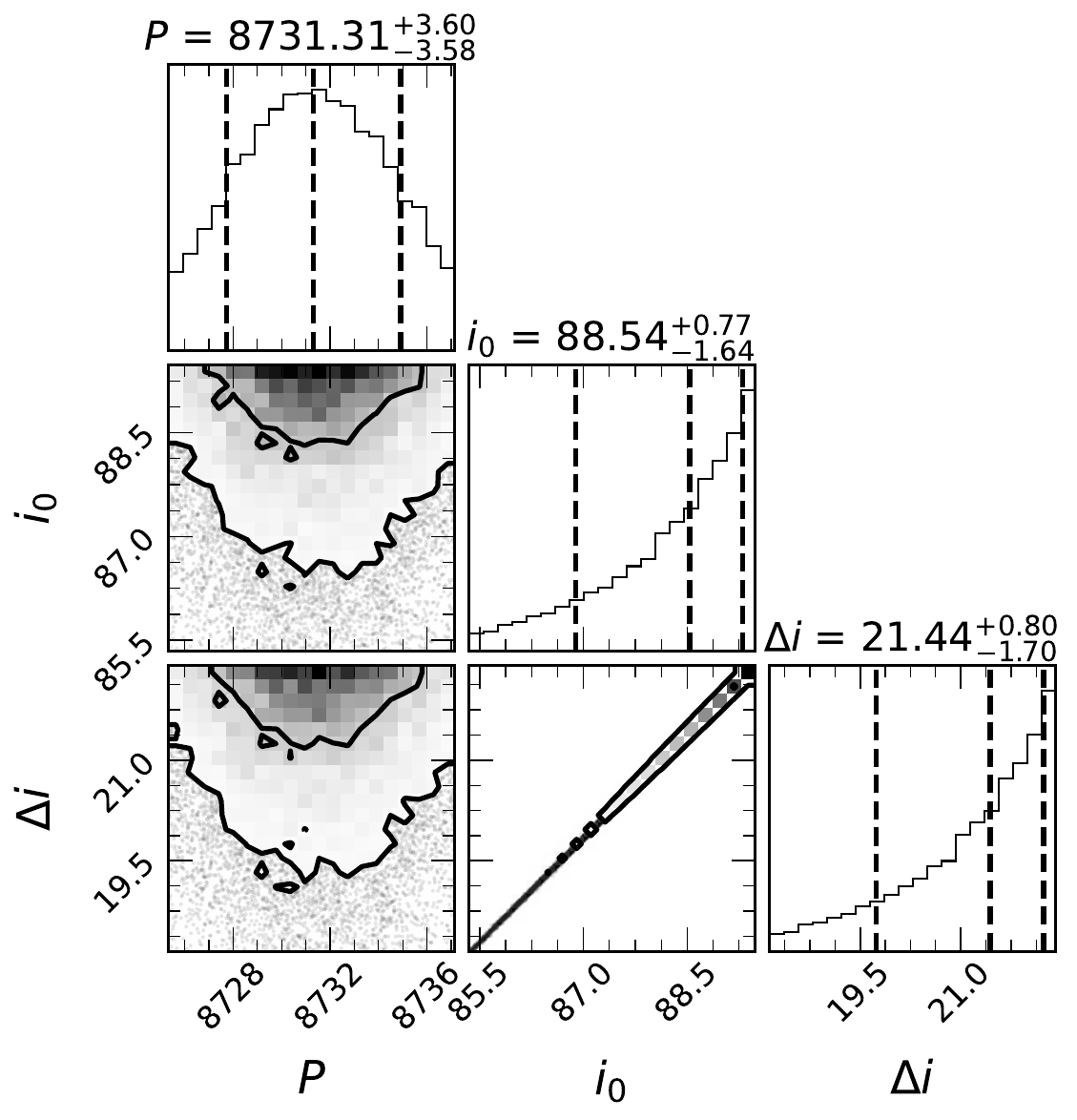}
    \includegraphics[width=0.49\textwidth]{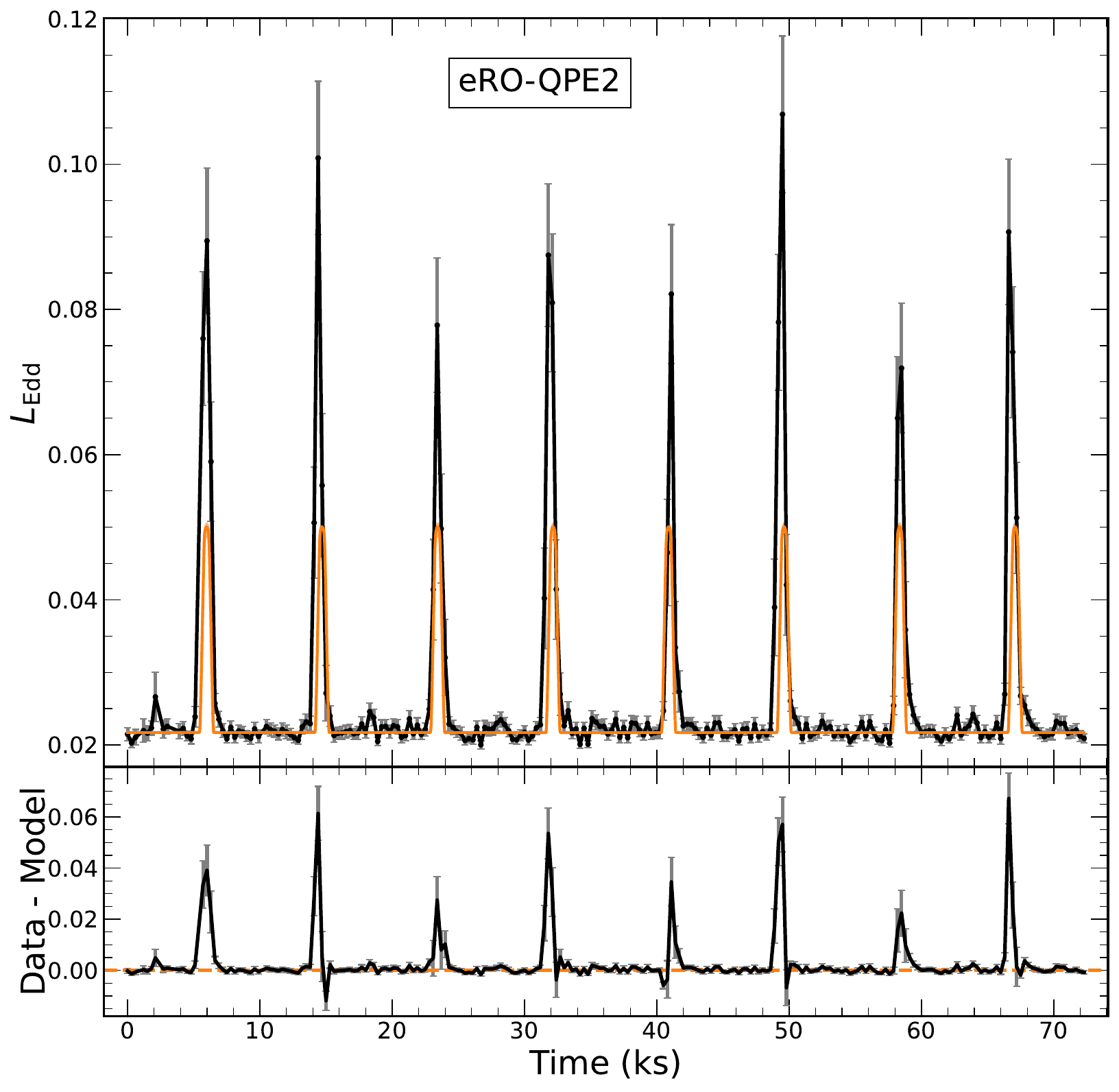}
    \includegraphics[width=0.49\textwidth]{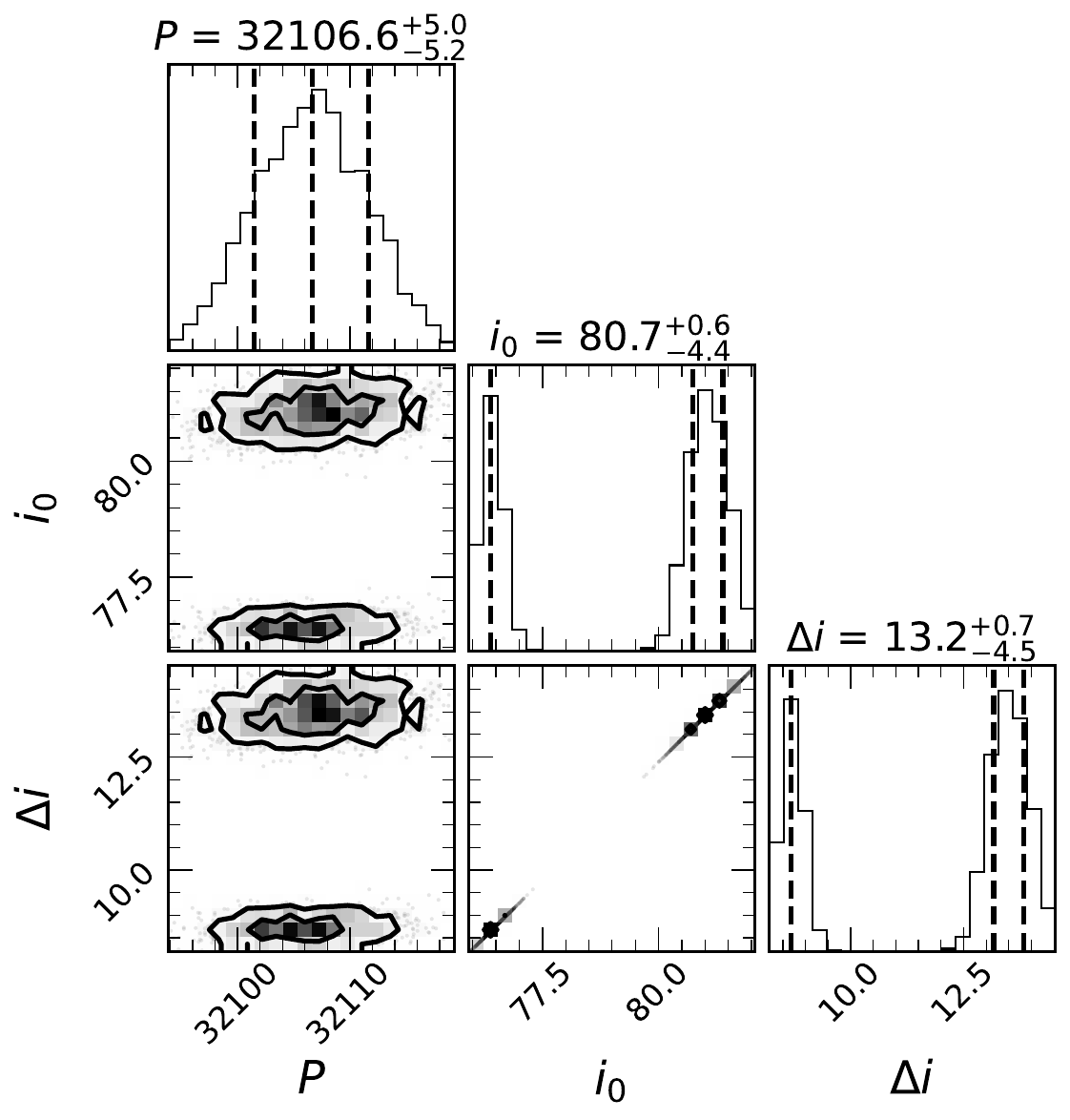}
    \includegraphics[width=0.49\textwidth]{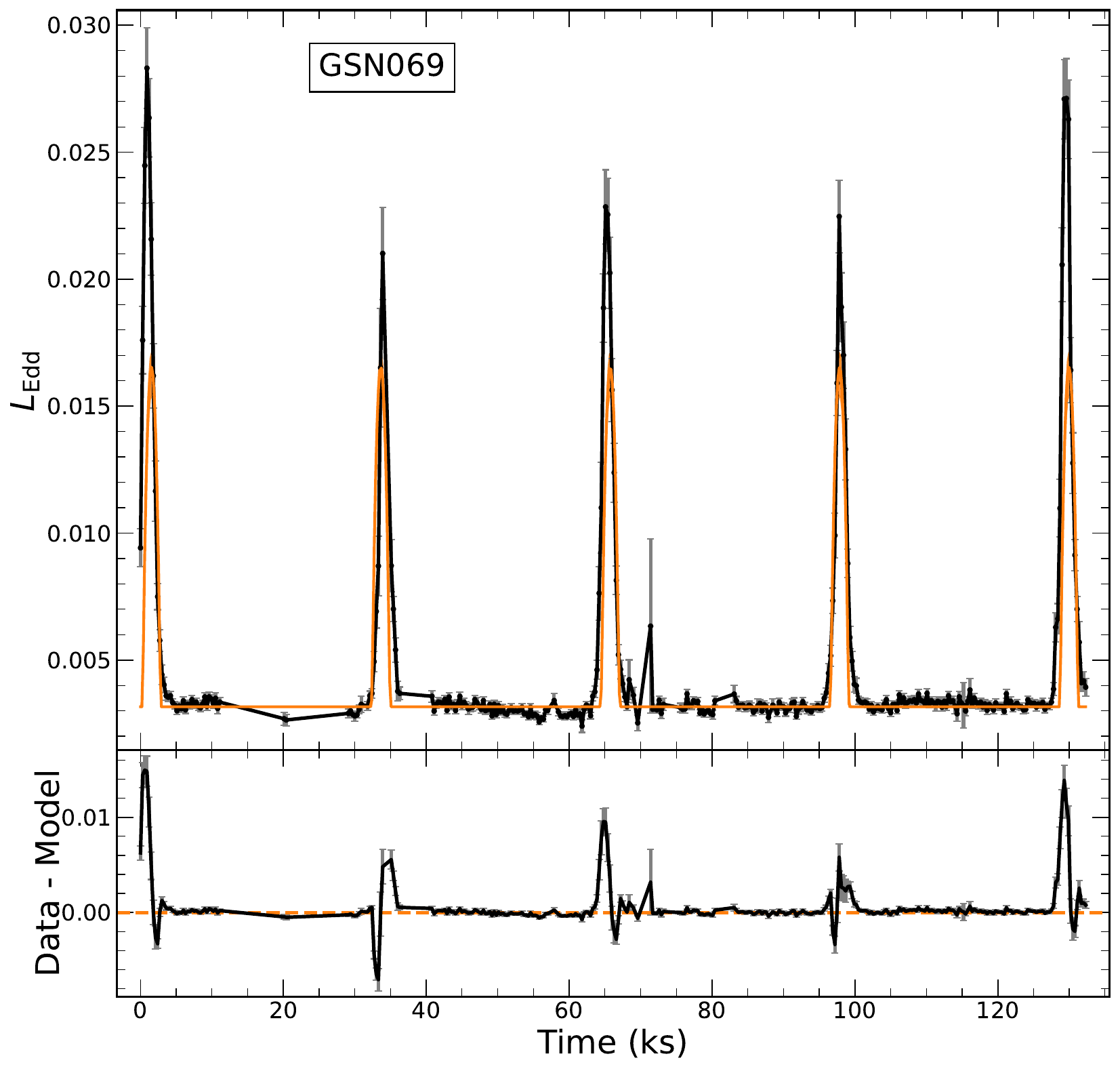}
    \caption{Constraints on the GR-RMHD-based precessing model fits to the lightcurves of eRO-QPE2 (top panels) and GSN069 (bottom panels). (Left) Posteriors with the 2-D 1 and 2$\sigma$ contours (39 and 86\% confidence levels respectively). The dashed lines on the histograms indicate the 16, 50 and 84\% percentiles (median $\pm 1\sigma$) respectively. Parameters $\phi, \dot{m}$ and $A$ have been omitted for clarity, as they are less relevant physically. (Right) Median model from the posteriors (orange solid line); the shaded areas indicate the 3 $\sigma$ confidence interval. Most of the solutions sampled from the posteriors are numerically equivalent, yielding unrealistically small errors. The lower panels show the residuals from the model whose parameters maximized the likelihood (or minimized $\chi^2$) and which are dominated by the events being somewhat brighter than the model, not being strictly periodic and not being the same flux in each event (as our simple model presently assumes).}
    \label{fig:ulxlc_posteriors}
\end{figure*}

\begin{table}
    \centering
    \begin{tabular}{cccc}
    \hline\hline
    \noalign{\smallskip}
    Parameter & Units & Value & Value \\
  & &   eRO-QPE2 & GSN069 \\ 
  \noalign{\smallskip}
   \hline
   \noalign{\smallskip}
         $P$ &  s &  8731$\pm4$ &  32107$\pm5$\\
         $\phi$ &  & 0.317$\pm0.003$ & 0.9519$\pm0.0003$ \\
         $i$ & $^\circ$ & 88.5$_{-1.6}^{+0.8}$ &  80.7$_{-4.4}^{+0.7}$ \\
         $\Delta i$ & $^\circ$ & 21.4$_{-1.7}^{+0.8}$ & 13.2$_{-4.5}^{+0.7}$ \\
         $\dot{m}$ & $\dot{M}_\mathrm{Edd}$ & 22$\pm$3 & 8$_{-3}^{+6}$ \\
          $A$ & 10$^{-3} L_\mathrm{Edd}$ & 21.71$\pm0.04$ & 3.13$\pm$0.01\\
        $\chi^{2}$/d.o.f. & & 611/230 & 2229/334 \\
          \noalign{\smallskip}
             \hline\hline
    \end{tabular}
    \caption{Constraints (median and 1$\sigma$ uncertainties) on the model parameters from fitting the lightcurves of eRO-QPE2 and GSN069. The parameters are, $P$: period in seconds, $\phi$: phase, $i$: inclination to the black hole spin axis, $\Delta i$: precession cone half-angle, $A$: offset, and $\dot{m}$: the Eddington-scaled accretion rate (at large radius). $\chi^2$ values are provided for the parameter combination which yielded the lowest $\chi^2$.
    }
    \label{tab:ulxlc_fits}
\end{table}

As shown in Figure~\ref{fig:ulxlc_posteriors}, the model can roughly reproduce the events. Although the best fits are statistically rejectable (Table~\ref{tab:ulxlc_fits}), we can see that the residuals are dominated by the model not quite reaching the required flux level (more clearly an issue for eRO-QPE2 than for GSN069) and the events not being strictly periodic or of the same peak flux, as noted by many authors. 
Nevertheless, both QPE lightcurves are described under a similar set of parameters; these imply a highly inclined system ($i > 70^\circ$) with a relatively small precessional half opening angle ($\Delta i \lesssim 20^\circ$) where, throughout all of the QPE phase, we view at high inclinations to the disc/wind (i.e. outside the half-opening angle of the funnel, $\sim$20\degr; Figure~\ref{fig:Ltheta}). 
We note that, according to our modelling, the fast rise is produced by a sharp change around $\sim67^\circ$ 
Such a sharp change is unlikely to be realistic, requiring 
a better understanding of the emission at large inclination and how this appears spectrally.

\section{Discussion}

Our Lense-Thirring precession model is able to explain several (although presently not all) key aspects of QPE phenomenology: 

\begin{itemize}
    \item The lack of substantial variability other than the events themselves can be explained by the large emitting area of the super-critical disc/wind. In our model, the estimated accretion rate for the QPEs lies between $\dot{m}$ = 1-20, implying light-crossing times and suppression of variability on timescales shorter than $\sim$100-1000s of seconds (depending on the details of the wind launching: \citealt{P07}). 
    \item The observed spectral evolution is also a reasonable match to models of super-Eddington accretion, with thermal spectra becoming hotter and brighter during the events. This behaviour would correspond to decreasing inclinations through the wind (e.g. \citealt{Middleton2015_spectraltiming, Dai2018}). However, there is a tension with the temperatures of the MC spectra extracted from the \texttt{HARMRAD} GR-RMHD simulations we have used (see Section 4.1.2). 
    \item The overall lightcurve shape can be approximated by a simple model of precession coupled with emission from a super-Eddington disc/wind. For the roughly periodic sources to which we have applied the model (GSN069 and eRO-QPE2), we inferred observer inclinations near to edge-on (noting that this is not the same as instantaneous inclination angle which is time dependent in our model). 
    \item At the inclinations implied by our modeling (see Table 2), we would infer a bolometric luminosity {\it in quiescence} of less than 10\% of Eddington which is broadly in agreement with observations (\citealt{Miniutti2023_GSNb}). 
    \item The changes in the recurrence time observed in QPEs (especially those which are strongly aperiodic, e.g. RXJ1301) can be explained within our model, as the precession period is highly sensitive to changes in the accretion rate (equation 1) via $r_{\rm o}$ which itself varies as either $\dot{m}$ (if $r_{\rm sph}$) or $\dot{m}^{3/2}$ (if $r_{\rm ph}$). As an example, should $\dot{m}$ at large radius change by $\Delta\dot{m}$ on timescales longer than the sound crossing time at $r_{\rm sph}$, then a relative change in precession period (i.e. $\Delta$P/P) of $\sim 3\Delta\dot{m}/\dot{m}$ is in principle possible (following the simplified equations of \citealt{Middleton2018, Middleton2019} for $\zeta = -1/2$, and assuming $\Delta\dot{m} < \dot{m}$). The observed extreme aperiodic nature of some events (e.g. eRO-QPE1 and RXJ1301) may then be due to large changes in $\dot{m}$ (e.g. the disc emptying) or some inherent instability in the fluid precession mechanism, e.g. when the outer radius of the precessing region approaches the limit set by the timescale for alignment (\citealt{Motta2018}).
    \item Should the disc be sufficiently compact e.g. if the infalling material has low angular momentum (as with a precursor TDE, see also \citealt{Miniutti2023_GSNa}) and is viewed at high inclinations (as we infer here), we could potentially see the emission from the wind cone from the underside of the flow (see Fig~\ref{fig:schem}) leading to a bright then faint event. 
    Taken to an extreme, observing both cones may offer one explanation for the long-short/bright-faint QPE behaviour seen in both GSN069 and eRO-QPE2 which is almost (although not-quite) anti-phase (\citealt{Arcordia2022_ero1behaviour}) or the dual peaks seen in RXJ1301 (\citealt{Franchini2023}, although in this case, the precessing central regions would need to be asymmetric about the disc plane). We note that we have have assumed a simple model in this first study where the recurrence time is from a single cone of emission. Whilst greater complexity is beyond the scope of this paper, we note that including emission from the undercone would lead to a longer recurrence time which could be matched with only a slightly higher accretion rate or a lower black hole spin value (see Equation 1).
    
    \item It has been observed that, in the case of GSN069, the QPE events appear to disappear when the source appears brighter in terms of bolometric luminosity (\citealt{Miniutti2023_GSNb}). This can potentially be accommodated within our model in a number of ways. The first option is that the accretion rate has increased such that the bolometric luminosity has increased (see Fig~\ref{fig:Ltheta}); the outer radius setting the disc precession will also increase and may become too large for stable precession to occur, leading to alignment (\citealt{Motta2018}). The second option is that the accretion rate has changed which has led to changes in the overall misalignment angle (requiring a change to the radial surface density profile, see \citealt{Fragile2007}) such that the flow is now viewed more face one, leading to higher luminosities and very low amplitude QPEs (following Fig~\ref{fig:Ltheta}). Finally, it may be that a change in accretion rate leads to a change in the structure of the disc and the accretion and sound crossing times, with precession then simply unable to occur (\citealt{Bollimpalli2024MNRAS}). All of these suggestions are speculative and need to be explored via simulations.  
    
\end{itemize}


\subsection{Predictions and challenges}

Our model makes a number of testable predictions and here we highlight both these as well as challenges to this interpretation of QPEs.\\

\subsubsection{Predictions}

One of the main predictions of our model is a clear dependence with accretion rate.
The lightcurve modelling of the inferred observed luminosities (which may underestimate the true luminosity to some extent given that advective discs have a flatter radial temperature profile), implies we are viewing the known QPE sources at inclinations which never enter the wind-cone. In the absence of any changes to the mean inclination angle, Fig~\ref{fig:tracks} would lead us to predict that a decrease in accretion rate should be accompanied by brighter QPEs with shorter recurrence times, and the flux ratio between weak and bright events (the latter being from the underside of the flow) could also increase. The opposite would happen with an increase in accretion rate (until the limit of stable precession is reached).

    

Given our model requires the presence of optically thick winds, we would expect the tentative spectral residuals seen at soft energies (Fig~\ref{fig:spectra1} and Fig~\ref{fig:spectra2}) be resolved into absorption (and likely some emission) features. As we infer the QPE sources to be seen at higher inclinations than ULXs, we would expect to detect strong winds with a slower velocity than seen in the latter, with the additional expectation that we should start to see higher projected velocities for the partially-ionised winds as our line-of-sight inclination decreases during an event. We note that, whilst these features appear absent from Fig~\ref{fig:GRsimspectra}, this is a resolution issue rather than indicating an absence of lines in the simulations. It is perhaps promising then that these residuals have been recently confirmed as indicating the presence of a slow moving, somewhat optically thick wind in the case of GSN069 (\citealt{Kosec2024}). 

Given the opening angle of the wind and X-ray luminosity shown in Fig~\ref{fig:Ltheta}, it is clear that, to avoid QPEs appearing strongly super-Eddington (even in the absence of a jet, which is a feature of magnetically dominated simulations), we would need to have observed all QPE sources thus far at $>45^{\circ}$, the probability of which is $\approx$70\%. This limiting angle may reduce further (thereby increasing the probability of observing a sub-Eddington QPE source) if misaligned discs are more advective in nature, which would allow lower luminosities to be reached at smaller inclinations. It is also clear that precession would only yield QPEs where the inclination changes yield major changes in luminosity, i.e. changes in inclination lying between 40-90$^{\circ}$ (as we have inferred for GSN069 and eRO-QPE2) or $\sim$0-20$^{\circ}$. The probability of observing the latter, brighter phase of QPEs is only $\sim$6\%.\\


\subsubsection{Challenges}

There remains an important observational issue our model is yet to address. QPE sources are mostly observed to follow a pattern of hysteresis (e.g. \citealt{Arcodia2024_ero3&4}), being somewhat harder and fainter on the rise and softer and brighter on the decay. This is shown for all the QPE sources in Fig~\ref{fig:tracks} based on our simple spectral modelling. Whilst it might not seem possible to create hysteresis paths from a precessing wind cone -- as inclination changes take a source forwards and backwards along the paths shown in Fig~\ref{fig:tracks} -- there are a number of additional factors which must be considered. From an observational perspective, the spectral model we have applied is too simplistic and does not describe a super-Eddington disc/wind -- we expect to tabulate a range of post-processed GR-RMHD spectra in the near future and will re-explore the fits to obtain a more accurate picture. However, should the present tension remain, another explanation would be needed. Whilst the asymmetry of some of the events would not naturally be explained by the simplest form of our model, the inclusion of relativistic effects -- as the disc/wind may be precessing at a few percent of c -- must lead to Doppler boosting of the emission, leading to overall asymmetry in the event and permitting the hard emission to peak before the soft (as observed in some sources: \citealt{Arcordia2022_ero1behaviour, Arcodia2024}). Obtaining a brighter decay phase seems harder to explain in this instance; one possibility to relieve this tension is if the disc and wind are precessing independently (as suggested in \citealt{Middleton2018} in the case of ULXs). For sources at only a few times Eddington, the precession periods are very similar for the disc and wind ($r_{\rm sph} \sim r_{\rm ph}$) but at higher rates these begin to differ substantially. Regardless of accretion rate, we should also expect a dynamical lag in the position of the outer photosphere on a timescale of order $r_{\rm ph}/v_{\rm wind}$. Such a lag or difference in period between the disc and wind (or the combination) could certainly introduce complex patterns of spectral-timing behaviour; we will explore these effects in future.

Another clear challenge our model must address is the temperature of the soft emission, where the observations are a factor $\sim 2$ (and even more so in the case of eRO-QPE2) higher than we would predict from our post-processing. This may be explained by the starting conditions for our MC calculations, which assume efficient Compton cooling with an input SED  set to be a blackbody peaking at 10$^6$K (this temperature being set by the GR-RMHD simulations with grey opacities); the starting conditions are a known issue for such post-processing and a correct energy balance should lead to harder spectra (see the discussion in {\citealt{Mills2023}}). The effect on the emergent spectrum will be especially pronounced at lower values of $\dot{m}$ where the wind is less optically thick. A further issue may arise from seeding our input SED only within the innermost radii, whilst in reality, emission must occur throughout the disc. We will explore the impact of both of these issues numerically in the future.

It has been noted by \cite{Miniutti2023_GSNb} that the intensity ratio of the pairs of QPEs in GSN069 is strongly correlated with the recurrence time (with the relative strengths inverting below some recurrence time), and poses a significant constraint on any model (see \citealt{Miniutti2023_GSNb} for discussion). In our LTP model, this change from leading to trailing event being brighter could potentially be explained by changes in accretion rate (which change the recurrence time) leading to a change in the mean misalignment angle about which the disc precesses (changing the relative notion of topside and underside). However, this remains a point of speculation we hope to address in future with numerical simulations.

\section{Conclusions}

We have extrapolated the model of Lense-Thirring precession of super-Eddington discs from ULXs (\citealt{Middleton2018, Middleton2019}) to QPEs. This is motivated by the growing association of QPEs with TDEs (\citealt{Miniutti2019Nature_2MASJ, Chakraborty2021, Nicholl2024, Bykov2024}) which, as pointed out by \cite{Quintin2023} in the case of AT2019vcb, is unlikely to be a coincidence given the low rates of TDEs (\citealt{vanVelzen2021}). 

Our model is simple but can describe several of the key characteristics of QPEs, both their temporal (their quasi-periodic to aperiodic nature, the rough shape and brightness of the events) and spectral properties. In all cases, we predict that these systems are accreting at one to a few times Eddington and are seen at high inclination angles (noting that this is not the same as the instantaneous angle to the wind cone). Our model makes clear predictions that powerful winds should be present and, should the residuals seen in the X-ray spectra be resolved into relativistically shifted atomic features (as they now have in GSN069: \citealt{Kosec2024}), this would be consistent with our model. 

There remain a number of challenges to this model, namely that, whilst the asymmetry of the events may be partly explained by relativistic effects or a lag between disc and outflow precession, the spectral hysteresis requires further consideration. This future work will benefit from further GR-RMHD simulations (covering a wider range of accretion rate and exploring misalignment), and post-processing with more physically consistent starting conditions up to higher inclinations. Finally, we note that our model would predict that QPEs should not need to arise necessarily from TDEs (though these are able to provide mass at super-Eddington rates) but from any super-Eddington accreting AGN as long as the Lense-Thirring torques can be efficiently communicated through the flow (i.e. when the sound crossing time is shorter than the precession timescale which, in turn, is shorter than the accretion timescale: \citealt{Bollimpalli2024MNRAS}). If such conditions are met, the predicted numbers of QPE-type events (which, at face-on inclinations might look more like the sinusoidal QPOs seen in Narrow line Seyfert AGN, e.g. \citealt{Gierlinksi2008, Lin2013, Ashton2021}, some of which have a strong spectral resemblance to QPE sources, e.g. \citealt{Terashima2012}) would therefore trace the high end of the AGN luminosity function. We note that observing such behaviour is sensitive to (at least) the accretion rate, SMBH mass and observer inclination.

 
\section*{Acknowledgements}

The authors thank the anonymous referee for their suggestions, and M. Guistini and N. Khan for helpful comments and discussion.
MM and AG acknowledge support via STFC Consolidated grant (ST/V001000/1). TK and LD acknowledge support from the National Natural Science Foundation of China and the Hong Kong Research Grants Council (HKU12122309, N\_HKU782/23, 27305119, 17305920, 17305523). AI acknowledges support from the Royal Society. G.M. was supported by grant PID2020-115325GB-C31 funded by MICIN/AEI/10.13039/501100011033. CP is supported by SEAWIND grant funded by NextGenerationEU.
\section*{Data Availability}

Data can be made accessible upon request. The code and data used in Section~\ref{sub:lightcurve_modelling} are available at \url{https://github.com/andresgur/qpes}.
 


\bibliographystyle{mnras.bst}
\bibliography{bibliography.bib}  

\appendix

\section{Spectral fitting} 

\begin{table}
\centering
\caption{Best fitting values and 1$\sigma$ errors for the parameters of interest, from spectral fitting to the various phases of the QPE events. The parameters are shown in chronological order through the events which -- in Fig {\ref{fig:spectra1}} and Fig {\ref{fig:spectra2}} follow the sequence black, blue, green, red, orange --  and are labelled according to the phases described in the corresponding papers (see text for details). nH values are provided in units of 10$^{22}$ cm$^{-2}$, kT$_{\rm disc}$ in keV, and L$_{\rm deabs}$ in 10$^{41}$ erg s$^{-1}$.}
 \begin{tabular}{||c c c ||}
 \hline
  \hline
 Source/Phase & Parameter & Value  \\
 \hline
 \texttt{} {\bf RXJ1301} & nH & 0.01$_{-0.00}^{+0.01}$ \\
    Rise 1           & kT$_{\rm disc}$ & 0.19$_{-0.01}^{+0.01}$ \\
               & L$_{\rm deabs}$ & 5.36$^{+0.16}_{-0.16}$\\
               \\
               & nH  & 0.01$_{-0.00}^{+0.01}$ \\
    Rise 2           & kT$_{\rm disc}$  & 0.19$_{-0.01}^{+0.01}$ \\
               & L$_{\rm deabs}$ & 13.70$^{+0.70}_{-0.25}$\\
               \\
               & nH  &  0.01$_{-0.00}^{+0.01}$\\
    Peak           & kT$_{\rm disc}$  & 0.17$_{-0.01}^{+0.01}$\\
               & L$_{\rm deabs}$ & 15.33$^{+0.49}_{-0.27}$\\
               \\
               & nH  &  0.01$_{-0.00}^{+0.01}$\\
    Decay 1           & kT$_{\rm disc}$  & 0.13$_{-0.01}^{+0.01}$ \\
               & L$_{\rm deabs}$  & 11.38$^{+0.94}_{-0.24}$\\
               \\
               & nH  & 0.01$_{-0.00}^{+0.02}$ \\
    Decay 2           & kT$_{\rm disc}$  & 0.10$_{-0.01}^{+0.01}$ \\
               & L$_{\rm deabs}$  & 5.80$^{+1.11}_{-0.20}$\\
               \hline
               &\(\chi^2/ \it{ dof}\) & 428.8/368 \\ 
               \hline
        {\bf GSN069} & nH &  0.02$_{-0.00}^{+0.01}$\\
  Rise 1             & kT$_{\rm disc}$ & 0.08$_{-0.01}^{+0.01}$\\
               & L$_{\rm deabs}$ & 4.61$^{+0.54}_{-0.18}$\\
               \\
               & nH & 0.02$_{-0.00}^{+0.01}$ \\
  Rise 2             & kT$_{\rm disc}$ & 0.13$_{-0.01}^{+0.01}$\\
               & L$_{\rm deabs}$ & 11.52$^{+0.50}_{-0.32}$\\
               \\
               & nH &  0.04$_{-0.01}^{+0.01}$\\
 Peak              & kT$_{\rm disc}$ & 0.12$_{-0.01}^{+0.01}$\\
               & L$_{\rm deabs}$ & 23.36$^{+1.68}_{-1.52}$\\
               \\
               & nH &  0.06$_{-0.01}^{+0.01}$\\
Decay 1            & kT$_{\rm disc}$ & 0.09$_{-0.01}^{+0.01}$\\
               & L$_{\rm deabs}$ & 20.23$^{+2.95}_{-2.46}$\\
               \\
               & nH &  0.03$_{-0.01}^{+0.01}$\\
 Decay 2              & kT$_{\rm disc}$ & 0.07$_{-0.01}^{+0.01}$\\
               & L$_{\rm deabs}$ & 6.51$^{+1.48}_{-0.96}$\\
                \hline
               & \(\chi^2/ \it{ dof}\) & 390.2/363\\ 
               \hline
        {\bf eRO-QPE1} & nH &  0.12$_{-0.02}^{+0.03}$\\
   Rise   & kT$_{\rm disc}$ & 0.12$_{-0.01}^{+0.01}$\\
               & L$_{\rm deabs}$ & 58.19$^{+13.14}_{-9.49}$\\
               \\
               & nH &  0.08$_{-0.01}^{+0.01}$\\
      Peak         & kT$_{\rm disc}$ & 0.12$_{-0.01}^{+0.01}$\\
               & L$_{\rm deabs}$ & 89.37$^{+8.60}_{-7.54}$\\
               \\
               & nH &  0.15$_{-0.01}^{+0.01}$\\
      Decay         & kT$_{\rm disc}$ & 0.08$_{-0.01}^{+0.01}$\\
               & L$_{\rm deabs}$ & 89.20$^{+13.40}_{-11.26}$\\
                \hline
               & \(\chi^2/ \it{ dof}\) & 276.1/256 \\ 
        \hline
 \hline
 \multicolumn{3}{c}{}
 \end{tabular}
 \label{table:a1}
\end{table}

\begin{table}
\centering
\caption{Best fitting values and 1$\sigma$ errors for the parameters of interest, from spectral fitting to the various phases of the QPE events. The parameters are shown in chronological order through the events which -- in Fig {\ref{fig:spectra1}} and Fig {\ref{fig:spectra2}} follow the sequence black, blue, green, red, orange --  and are labelled according to the phases described in the corresponding papers (see text for details). nH values are provided in units of 10$^{22}$ cm$^{-2}$, kT$_{\rm disc}$ in keV, and L$_{\rm deabs}$ in 10$^{41}$ erg s$^{-1}$.}
 \begin{tabular}{||c c c ||}
  \hline
 \hline
 Source/Phase & Parameter & Value  \\
 \hline
  {\bf eRO-QPE2} & nH &  0.19$_{-0.08}^{+0.09}$\\
  Rise      & kT$_{\rm disc}$ & 0.24$_{-0.04}^{+0.04}$\\
               & L$_{\rm deabs}$ & 3.08$^{+1.35}_{-0.81}$\\
               \\
               & nH &  0.25$_{-0.04}^{+0.04}$\\
   Peak            & kT$_{\rm disc}$ & 0.26$_{-0.01}^{+0.01}$\\
               & L$_{\rm deabs}$ & 8.66$^{+1.32}_{-1.08}$\\
               \\
               & nH &  0.17$_{-0.04}^{+0.04}$\\
    Decay           & kT$_{\rm disc}$ & 0.17$_{-0.01}^{+0.01}$\\
               & L$_{\rm deabs}$ & 3.19$^{+0.89}_{-0.63}$ \\
                \hline
               & \(\chi^2/ \it{ dof}\) &  82.2/80\\ 
               \hline
       {\bf eRO-QPE3} & nH &  0.06$_{-0.04}^{+0.13}$\\
   Rise 2      & kT$_{\rm disc}$ & 0.15$_{-0.03}^{+0.03}$\\
               & L$_{\rm deabs}$ & 1.75$^{+2.56}_{-0.55}$\\
               \\
               & nH &  0.2$_{-0.1}^{+0.2}$\\
   Peak            & kT$_{\rm disc}$ & 0.10$_{-0.03}^{+0.04}$\\
               & L$_{\rm deabs}$ & 8.27$^{+37.87}_{-5.57}$\\
               \\
               & nH & 0.02$_{-0.02}^{+0.16}$ \\
    Decay 1           & kT$_{\rm disc}$ & 0.21$_{-0.08}^{+0.08}$\\
               & L$_{\rm deabs}$ & 0.75$^{+1.38}_{-0.12}$\\
               \\
               & nH &  0.5$_{-0.4}^{+0.8}$\\
    Decay 2           & kT$_{\rm disc}$ & 0.06$_{-0.03}^{+0.04}$\\
               & L$_{\rm deabs}$ & 27.81$^{+1182.87}_{-26.58}$\\
                \hline
               & \(\chi^2/ \it{ dof}\) & 15.7/12 \\ 
               \hline
        {\bf eRO-QPE4} & nH &  0.06$_{-0.00}^{+0.01}$\\      
  Rise 2      &  kT$_{\rm disc}$ & 0.15$_{-0.01}^{+0.01}$\\
               & L$_{\rm deabs}$ & 49.48$^{+3.60}_{-2.13}$\\
               \\
               & nH & 0.08$_{-0.01}^{+0.01}$ \\
     Peak         & kT$_{\rm disc}$ & 0.15$_{-0.01}^{+0.01}$\\
               & L$_{\rm deabs}$ & 110.26$^{+22.27}_{-16.44}$\\
               \\
               & nH & 0.13$_{-0.03}^{+0.03}$ \\
   Decay 1            & kT$_{\rm disc}$ & 0.09$_{-0.01}^{+0.01}$\\
               & L$_{\rm deabs}$ & 111.49$^{+73.88}_{-39.52}$\\
               \\
               & nH & 0.06$_{-0.00}^{+0.02}$\\
    Decay 2           & kT$_{\rm disc}$ & 0.07$_{-0.01}^{+0.01}$\\
               & L$_{\rm deabs}$ & 8.73$^{+6.28}_{-1.44}$\\
                \hline
               & \(\chi^2/ \it{ dof}\) & 165.2/161\\ 
     
 \hline
 \hline
 \multicolumn{3}{c}{}
 \end{tabular}
 \label{table:a2}
\end{table}

\begin{table}
\centering
\caption{Best fitting values and 1$\sigma$ errors for the parameters of interest, from spectral fitting to the various phases of the QPE events for 2MASJ. The parameters are shown in chronological order through the events which -- in Fig {\ref{fig:spectra2}} for 2MASJ -- follow the sequence black, blue, green, and are labelled according to the phases described in the corresponding papers (see text for details). nH values are provided in units of 10$^{22}$ cm$^{-2}$, kT$_{\rm disc}$ in keV, and L$_{\rm deabs}$ in 10$^{41}$ erg s$^{-1}$.}
 \begin{tabular}{||c c c ||}
  \hline
 \hline
 Source/Phase & Parameter & Value  \\
 \hline
{\bf 2MASJ} & nH & 0.17$_{-0.07}^{+0.09}$ \\
Rise       & kT$_{\rm disc}$ & 0.14$_{-0.02}^{+0.02}$\\
               & L$_{\rm deabs}$ & 14.20$^{+12.61}_{-5.42}$\\
               \\
               & nH & 0.03$_{-0.00}^{+0.06}$ \\
 Peak              & kT$_{\rm disc}$ & 0.24$_{-0.05}^{+0.03}$\\
               & L$_{\rm deabs}$ & 6.30$^{+2.53}_{-0.50}$\\
\\
               & nH & 0.3$_{-0.2}^{+0.2}$ \\
  Decay             & kT$_{\rm disc}$ & 0.08$_{-0.02}^{+0.03}$\\
               & L$_{\rm deabs}$ & 59.02$^{+339.79}_{-43.21}$\\
                \hline
         & \(\chi^2/ \it{ dof}\) & 25.7/20 \\ 
 \hline
 \hline
 \multicolumn{3}{c}{}
 \end{tabular}
 \label{table:a3}
\end{table}




\bsp	
\label{lastpage}
\end{document}